\documentclass[amssymb,amsmath,aps,showpacs,floatfix,nofootinbib,12pt]{revtex4}
\usepackage{amssymb}
\usepackage{graphicx}
\usepackage{color}
\usepackage{latexsym}

\begin{document}

\voffset 1.25cm

\title{Gamma rays and neutrinos from dark matter annihilation in
galaxy clusters}

\author{Qiang Yuan$^{1}$, Peng-Fei Yin$^{2}$, Xiao-Jun Bi$^{1,3}$,
Xin-Min Zhang$^{4}$,  and Shou-Hua Zhu$^{2,3}$}

\affiliation{$^{1}$ Key Laboratory of Particle Astrophysics,
Institute of High Energy Physics, Chinese
Academy of Sciences, Beijing 100049, P. R. China \\
$^{2}$ Institute of Theoretical Physics \& State Key Laboratory of
Nuclear Physics and Technology, Peking University,
Beijing 100871, P.R. China \\
$^{3}$ Center for High Energy Physics,
Peking University, Beijing 100871, P.R. China \\
$^{4}$ Theoretical Division, Institute of High Energy Physics,
Chinese Academy of Sciences, Beijing 100049, P. R. China }

\date{\today}

\begin{abstract}

The $\gamma$-ray and neutrino emissions from dark matter (DM)
annihilation in galaxy clusters are studied. After about one year
operation of Fermi-LAT, several nearby clusters are reported with
stringent upper limits of GeV $\gamma$-ray emission. We use the
Fermi-LAT upper limits of these clusters to constrain the DM model
parameters. We find that the DM model distributed with
substructures predicted in cold DM (CDM) scenario is strongly
constrained by Fermi-LAT $\gamma$-ray data. Especially for the
leptonic annihilation scenario which may account for the $e^{\pm}$
excesses discovered by PAMELA/Fermi-LAT/HESS, the constraint on
the minimum mass of substructures is of the level $10^2-10^3$
M$_{\odot}$, which is much larger than that expected in CDM
picture, but is consistent with a warm DM scenario. We further
investigate the sensitivity of neutrino detections of the clusters
by IceCube. It is found that neutrino detection is much more
difficult than $\gamma$-rays. Only for very heavy DM ($\sim 10$
TeV) together with a considerable branching ratio to line
neutrinos the neutrino sensitivity is comparable with that of
$\gamma$-rays.

\end{abstract}

\pacs{95.35.+d,95.85.Pw,95.85.Ry,98.65.Cw}

\maketitle

\newpage

\section{Introduction}

The existence of dark matter (DM) has been established by many
astrophysical observations, but the nature of DM paticle is still
unclear. Among the large amount of candidates proposed in many
theories of new physics, the weakly interacting massive particle
(WIMP) is the most popular and attractive one
\cite{1996PhR...267..195J,2005PhR...405..279B}. The mass of WIMP
is generally from a few GeV to TeV, and the interaction strength
is of the weak scale, which can give the right relic density of
DM. In this scenario, the weak interaction of DM particles would
produce observable standard model particles, such as charged
anti-matter particles, photons and neutrinos. Investigating
such particles from the cosmic rays (CRs) is the task of DM
indirect detection.

The recently reported new signatures of CR positrons, antiprotons
and electrons by PAMELA
\cite{2009Natur.458..607A,2009PhRvL.102e1101A}, ATIC
\cite{2008Natur.456..362C}, HESS \cite{2008PhRvL.101z1104A,
2009A&A...508..561A} and Fermi-LAT \cite{2009PhRvL.102r1101A} have
stimulated great interests and extensive studies of the DM indirect
searches. The DM scenario with mass $O$(TeV), leptonic
annihilation/decay final states and a high annihilation/decay rate
can well explain the observational data (e.g.,
\cite{2009NuPhB.813....1C,2009PhRvD..79b3512Y}). Furthermore, more
quantitative constraints on the DM model parameters can be derived
through a global fitting method
\cite{2010PhRvD..81b3516L,2009arXiv0911.1002L}.

Regardless of detailed models of DM to explain the data, it is
essential to find observable signals to test the models. Since the
charged particles will gyrate in the magnetic field and lose most of
the source information, it is difficult to test the DM models using
only the data of charged CRs. Gamma-rays and neutrinos seem to be
very good probes. There are several advantages of using $\gamma$-ray
photons and neutrinos to investigate the DM models. Firstly, photons
and neutrinos propagate along straight line and can trace back to
the source sites where the DM annihilation/decay takes place.
Secondly there is little interaction during the propagation and most
of the primary source information hold. Thirdly the effective volume
of which photons and neutrinos can reach is much larger than that of
charged particles, e.g., from the Milky Way to extragalactic space,
and even the early Universe. It has been shown in some works that
$\gamma$-rays and neutrinos can be powerful tools to test the DM
scenarios explaining the CR lepton data (e.g.,
\cite{2009JCAP...03..009B,2009PhRvD..80b3007Z,2009PhRvD..79h1303B,
2009arXiv0908.1236Z,2009arXiv0912.0663C,2010JCAP...03..014P,
2009arXiv0912.4504Z,2009arXiv0908.4317C},
\cite{2009PhRvD..79d3516H,2009PhRvD..79f3522L,2009arXiv0905.4764S,
2010PhRvD..81a6006B,2010PhRvD..81d3508M,2010PhRvD..81h3506S,
2010JCAP...04..017C}).

There are many sites proposed to be good candidates for the search
of $\gamma$-rays and neutrinos from DM, such as the Galactic center
\cite{2009JCAP...03..009B,2009PhRvD..80b3007Z,2009PhRvD..79h1303B},
Galactic halo \cite{2009arXiv0908.1236Z,2009ApJ...699L..59B,
2010JCAP...03..014P}, satellite galaxies or substructures
\cite{2009PhRvD..80b3506E,2009MNRAS.399.2033P,
2009Sci...325..970K,2009arXiv0908.0195P}, the extragalactic space
\cite{2009PhRvD..80b3517K,2009JCAP...07..020P,2010MNRAS.405..593Z}
and the emissions at the early Universe
\cite{2009A&A...505..999H,2009JCAP...10..009C,2009arXiv0912.2504Y}.
As the largest gravitational bounding system in the Universe, galaxy
clusters may also be useful for DM indirect searches
\cite{2009PhRvD..80b3005J}. Pinzke et al. investigated the
$\gamma$-ray emission from nearby clusters and used EGRET upper
limits to set constraints on the DM model parameters
\citep{2009PhRvL.103r1302P}. They found that if the DM annihilation
was responsible for the electron/positron excesses and the
luminosity-mass distribution of DM substructures in clusters
followed the extrapolation of numerical simulation results, the
minimum mass of DM subhalos should be larger than $10^{-2}$
M$_{\odot}$ in order not to exceed the EGRET limits. This is a
useful way to study the particle nature of DM through structures.

After more than one year's operation, Fermi-LAT reported some
results about the $\gamma$-ray emission from galaxy clusters
\cite{Fermi-LAT:cluster}. Non detection of significant $\gamma$-ray
emission from galaxy clusters was reported except for Perseus
cluster, in which the emission from the central galaxy NGC 1257 was
discovered \cite{2009ApJ...699...31A}. The upper limits given by
Fermi-LAT are lower by more than one order of magnitude than that
given by EGRET. It can be expected that the new results from
Fermi-LAT will set much stronger constraints on the DM models.
In Ref. \cite{Fermi-LAT:cluster} the constraints on DM mass and
annihilation cross section were presented assuming $\mu^+\mu^-$ and
$b\bar{b}$ channels. In this work we will also use the Fermi-LAT upper
limits to constrain DM model parameters. Different from Ref.
\cite{Fermi-LAT:cluster}, we will pay more attention on the implication
of DM structure properties such as the minimal mass of subhalo
$M_{\rm min}$, which would be important for understanding the nature
of DM particle. This is one of the motivations of this study.

Another motivation of this work is the neutrino emission. Neutrinos
can be served as an independent diagnostic of DM indirect searches
besides photons. It has been shown that the measured atmospheric
neutrino background can set effective constraints on the DM
annihilation cross section
\cite{2007PhRvL..99w1301B,2007PhRvD..76l3506Y}. There are no high
energy astrophysical neutrinos being detected currently, so it is
valuable to explore the sensitivity of the forthcoming neutrino
detectors to the neutrino signals from the DM annihilation. Due to
the very weak interaction cross section between neutrinos and
matter, we generally need large detector volume. The ongoing
experiment IceCube has an effective volume $\sim$km$^3$, which would
give unprecedented sensitivity for the neutrino detection up to very
high energies. The detectability of neutrino signals from DM
annihilation in galaxy clusters by e.g. IceCube, will be discussed
in this work.

This paper is organized as follows. In Sec. II, we discuss the
$\gamma$-ray emission from several galaxy clusters, and employ the
recent Fermi-LAT limits of these clusers to constrain the DM model
parameters. In Sec. III, we discuss the detectability of neutrino
emission from the galaxy clusters by the neutrino detectors. The
last section is our conclusions and discussions.

\section{Gamma rays from galaxy clusters}

\subsection{Cluster sample}

It is known that the objects with high masses and small distances will
be very efficient for the DM searches. Therefore nearby massive clusters
are the first choice of study. Here we adopt a sample of $6$ clusters
with redshift from $0.0031$ to $0.0231$ (corresponding to distance from
$13$ to $100$ Mpc for a standard $\Lambda$CDM cosmology), which are
reported with flux upper limits by Fermi-LAT. The basic parameters
of these clusters are compiled in Table \ref{table:sample}.

The flux of DM annihilation from cluster is generally scaled with
$M_{200}^{\alpha}/d^2$, where $\alpha$ depends on the concentration-mass
relation and DM profile of the halo. In this work we will assume NFW
profile for the halo of cluster. The final results are expected not
sensitively dependent on the profile since most of the cluster lies in
the angular window of Fermi-LAT ($3.5^{\circ}$ for 100 MeV,
\cite{2009ApJ...697.1071A}). As a benchmark configuration, we adopt the
concentration-mass relation fitted from X-ray observations
\cite{2007ApJ...664..123B}
\begin{equation}
c_{\rm vir}=\frac{9.0}{1+z}\times\left(\frac{M_{\rm vir}}{10^{14}h^{-1}
{\rm M}_{\odot}}\right)^{-0.172}.
\end{equation}
After correcting the definition of virial overdensity in Ref.
\cite{2007ApJ...664..123B} ($\Delta\approx 100$) to $\Delta=200$
we have \cite{2003ApJ...584..702H}
\begin{equation}
c_{200}=\frac{6.9}{1+z}\times\left(\frac{M_{200}}{10^{14}
{\rm M}_{\odot}}\right)^{-0.178}.
\end{equation}
For this concentration-mass relation we find $\alpha\approx 0.65$.
Comparing the quantity $M_{200}^{\alpha}/d^2$ among these clusters,
we find that DM signals from clusters NGC 4636, M49 and Fornax are
of the same level, and are several times larger than the rest three
clusters. In the following we will see that these three clusters
will indeed give stronger constraints on the DM models.

\begin{table}[htb]
\centering
\caption{Cluster sample}
\begin{tabular}{cccccc}
\hline \hline
 Name & $z$\footnotemark[1] & R.A.\footnotemark[1] & Dec.\footnotemark[1] & $M_{200}$($10^{14}$M$_{\odot}$)\footnotemark[2] & $r_{200}$(Mpc)\footnotemark[2] \\
\hline
NGC 4636 & $0.0031$ & $12^h43^m$ & $2^{\circ}41'$ & $0.25$ & $0.60$\\
M49 & $0.0033$ & $12^h30^m$ & $8^{\circ}00'$ & $0.46$ & $0.73$\\
Fornax & $0.0046$ & $03^h39^m$ & $-35^{\circ}27'$ & $1.00$ & $0.95$\\
Centaurus & $0.0114$ & $12^h49^m$ & $-41^{\circ}18'$ & $2.66$ & $1.32$\\
AWM 7 & $0.0172$ & $02^h55^m$ & $41^{\circ}35'$ & $4.28$ & $1.54$\\
Coma & $0.0231$ & $13^h00^m$ & $27^{\circ}59'$ & $13.65$ & $2.27$\\
  \hline
  \hline
\end{tabular}
\footnotetext[1]{Redshift and coordinates are adopted from NASA/IPAC
Extragalactic Database, http://nedwww.ipac.caltech.edu/}
\footnotetext[2]{Virial mass and radius parameters are taken from
Ref. \cite{2002ApJ...567..716R}.}
\label{table:sample}
\end{table}

\subsection{Gamma-ray emission from DM distribution in clusters}

There are generally two kinds of $\gamma$-ray emission from DM
annihilation: one is produced directly from the annihilation final
state particles which is called {\it primary} emission (such as
the $\gamma$ rays by $\pi^0$ decay after hadronization or emission
directly from final charged leptons), and the other is produced
through interactions of final state particles with external medium
or radiation field such as the inverse Compton (IC) radiation
which is called {\it secondary} emission hereafter. The primary
$\gamma$-ray flux observed on the Earth from DM annihilation in a
galaxy cluster can be expressed as
\begin{equation}
\phi^{pri}=\frac{\langle\sigma v\rangle}{2m_{\chi}^2}\frac{{\rm d}N}
{{\rm d}E}\times \frac{\int \rho^2(r) {\rm d}V}{4\pi d_L^2},
\label{phi}
\end{equation}
where $m_\chi$ is the mass of DM particle, $\langle\sigma
v\rangle$ is the annihilation cross section of DM, $\frac{{\rm
d}N}{{\rm d}E}$ is the yield spectrum of $\gamma$-rays per
annihilation which is simulated using PYTHIA
\cite{2006JHEP...05..026S}, $d_L$ is the luminosity distance of
the cluster, $\rho(r)$ is the density distribution of DM inside
the cluster with $r$ the distance from the cluster center. All of the
cluster is taken into account in the integral since the analysis of
Fermi-LAT was done in a $10$ degree radius of each cluster
\cite{Fermi-LAT:cluster}, which is large enough to contain the whole
cluster halo. For the smooth halo we assume the density distribution
to be NFW profile \cite{1997ApJ...490..493N}
\begin{equation}
\rho_{\rm sm}(r)=\frac{\rho_s}{(r/r_s)(1+r/r_s)^2},
\label{nfw}
\end{equation}
where parameters $r_s$ and $\rho_s$ can be determined by the
concentration-mass relation and normalization of total mass.

Since there are substructures in the clusters, such as galaxy
groups and galaxies, we have to take these into account. We will
see later that the existence of substructures enhances the
annihilation luminosity of DM and is the main reason that affects
the $\gamma$ flux. To take into account the effect of
substructures, we replace $\rho^2$ in Eq. (\ref{phi}) with
$\rho^2_{\rm tot}\equiv\rho_{\rm sm}^2 +\langle\rho_{\rm
sub}^2\rangle$, where the average density square of substructures
reads
\begin{equation}
\langle\rho_{\rm sub}^2(r)\rangle=\int{\rm d}M\frac{{\rm d}N}
{{\rm d}V{\rm d}M}\times L(M),
\label{sub}
\end{equation}
in which $\frac{{\rm d}N}{{\rm d}V{\rm d}M}$ is the number density of
subhalos in mass bin ${\rm d}M$, $L(M)=\int_{V_{\rm sub}}\rho_{\rm sub}^2
{\rm d}V'$ is the intrinsic annihilation luminosity of a subhalo with mass
$M$. In this work we will employ the results from recent high resolution
simulation, {\it Aquarius} \cite{2008Natur.456...73S,2008MNRAS.391.1685S}
to treat the subhalos. Because the concentration and density profile of
subhalos are very complicated inside the host halo, the detailed
computation using Eq. (\ref{sub}) is difficult. Thus we directly
adopt the counted results of luminosity distribution from the
simulation\footnote{This relation is different from that given in Ref.
\cite{2008Natur.456...73S}, where $L(>M)\approx M^{-0.226}$ was found.
According to this fit we have $L(>M)\approx \left(M^{-0.16}-
M_{\rm max}^{-0.16}\right)$, which gives similar behavior as that in
Ref. \cite{2008Natur.456...73S} in the large (resolved) mass range of
the simulation, but is different when extrapolating to low (unresolved)
mass range.}
\cite{Yuan:neutrino}
\begin{equation}
\frac{{\rm d}{\mathcal L}}{{\rm d}M}(r,M)\equiv\frac{{\rm d}N}{{\rm d}V
{\rm d}M}\times L(M)\propto \left(\frac{r}{0.2r_{200}}\right)^{-0.1}
\left(1+\frac{r}{0.2r_{200}}\right)^{-2.9}\times M^{-1.16}.
\label{lumin_sub}
\end{equation}
Similar with Ref. \cite{2009PhRvL.103r1302P} we adopt a scale between
the Milky Way like halo given in {\it Aquarius} simulation and the case
of clusters, i.e., the ratio of ${\mathcal L}_{\rm sub}/{\mathcal L}_
{\rm sm}$ keeps unchanged whatever the mass is. The maximum mass of
subhalos found in simulation is about $0.01M_{\rm host}$. But the
minimum mass is not well known due to the limit of resolution of the
numerical simulation. From the observational point of view, we have
observed DM halos with mass $\sim 10^7$ M$_{\odot}$, e.g. dwarf
galaxies. While the study of free streaming of cold DM (CDM) particles
indicates a minimum halo mass down to $\sim 10^{-7}$ M$_{\odot}$
\cite{2001PhRvD..64h3507H}. In this work we leave $M_{\rm min}$ to
be a free parameter and investigate the effects of $M_{\rm min}$ on
the DM signals.

Besides the primary $\gamma$-ray emission, there is also secondary
production of $\gamma$-ray photons through the IC scatterings between
the DM induced electrons/positrons and the cosmic microwave background
(CMB) field. For the calculation of the fluxes of secondary IC emission
please see the Ref. \cite{2009arXiv0908.1236Z}. Note that when
calculating the energy loss
rate of electrons/positrons, both the IC loss induced by scattering with CMB
photons and the synchrotron loss in the magnetic field in clusters are
considered. The average value of magnetic field strength is assumed to
be $\sim 1$ $\mu$G \cite{1998APh.....9..227C}.

\begin{figure}[!htb]
\begin{center}
\includegraphics[width=0.47\columnwidth]{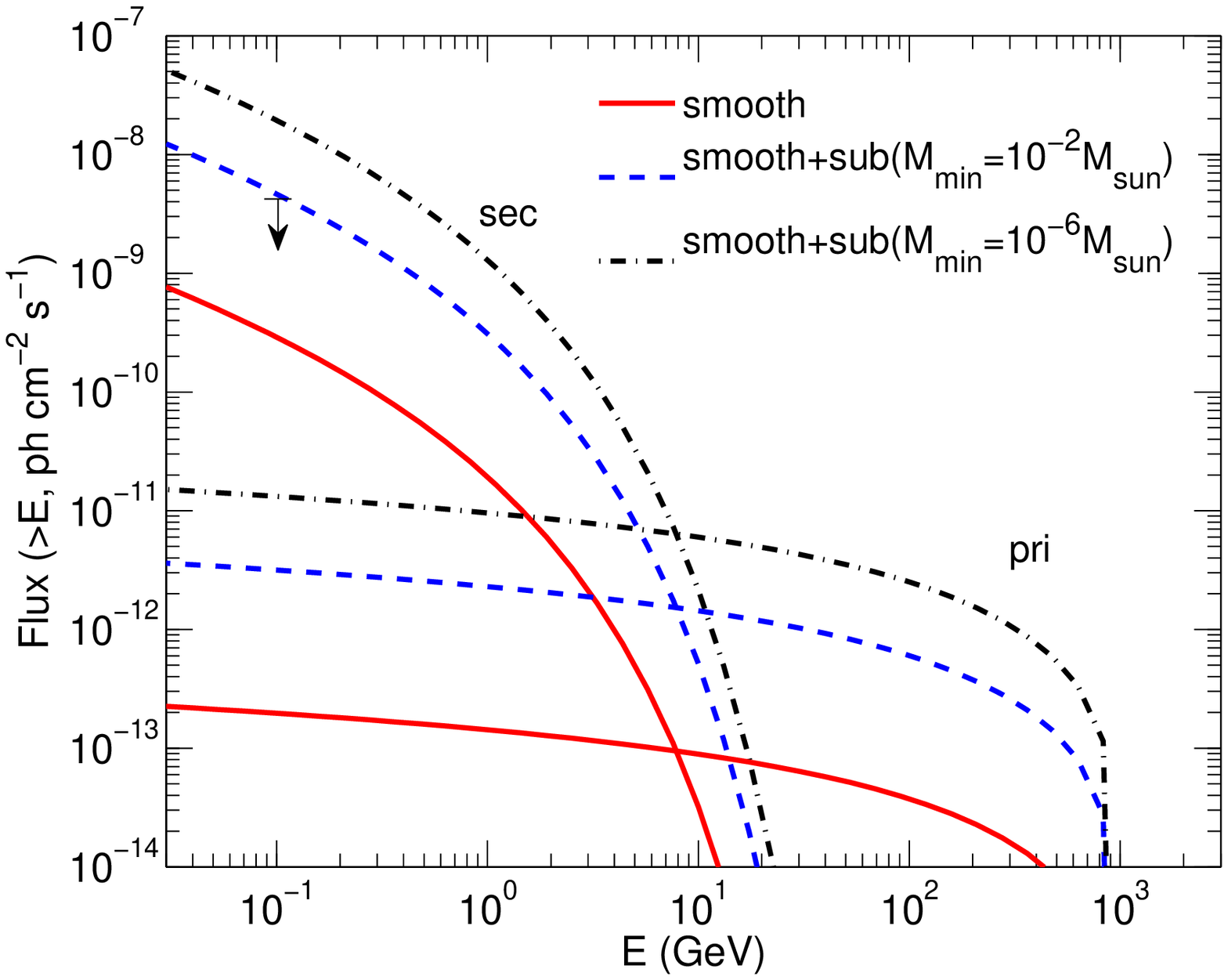}
\includegraphics[width=0.47\columnwidth]{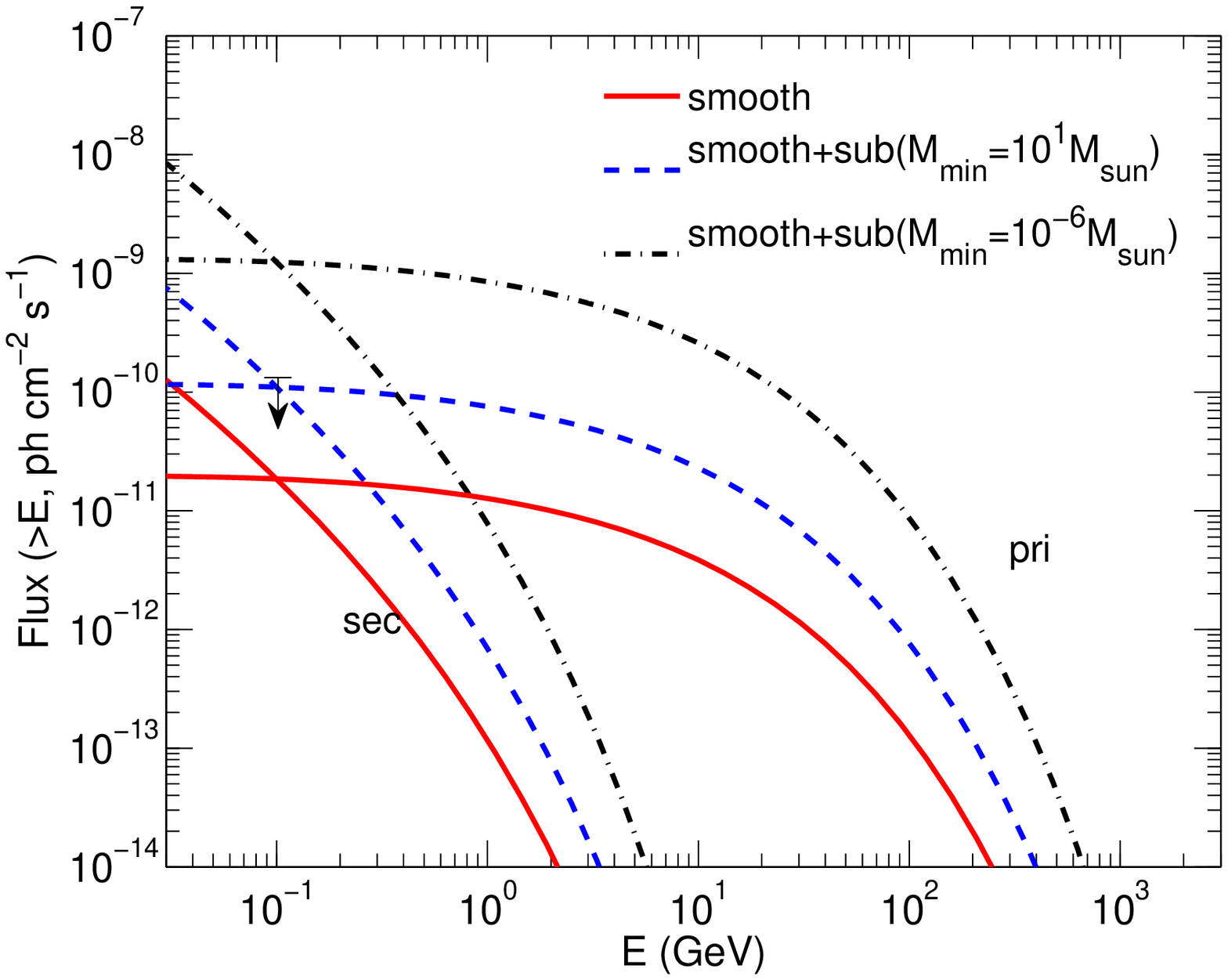}
\caption{Integral spectra of the IC and FSR components from DM
annihilation in Fornax cluster. The left panel is for $\mu^+\mu^-$
channel, and the right panel is for $b\bar{b}$ channel respectively.
Also shown are the $95\%$ upper limits (arrows) of Fermi-LAT 1 yr
observations. See the text for details.
\label{fig:spec}}
\end{center}
\end{figure}

For illustration we show the integral spectra of $\gamma$-rays
from DM annihilation in Fornax cluster in Fig. \ref{fig:spec}.
Here we adopt a sample DM model with $m_\chi=1$ TeV,
$\langle\sigma v\rangle=10^{-23}$ cm$^3$ s$^{-1}$, and the
annihilation channels with $\mu^+\mu^-$ (left) and $b\bar{b}$
(right) respectively. The {\it primary} and {\it secondary}
components are shown separately. In each group we show three
curves which represent the smooth halo contribution, the total
emission with subhalos down to two different $M_{\rm min}$. The
$95\%$ confidence level upper limits from Fermi-LAT are shown by
arrows. It is shown that in the energy range interested here,
i.e. $0.1-10$ GeV, the {\it secondary} radiation from IC is dominant
for $\mu^+\mu^-$ channel, while for $b\bar{b}$ channel the
{\it primary} contribution is dominant. This is because for $\mu^+\mu^-$
final state the spectra of photons and electrons are relatively hard,
and the {\it secondary} produced photons through IC can just lie in
the interested energy range. For $b\bar{b}$ final state the energies
of photons and electrons from the hadronic cascade are generally much
lower than the mass of DM, so the IC component dominates at even lower
energies ($<0.1$ GeV). This conclusion will always hold for $m_\chi$
ranging from $100$ GeV to $10$ TeV for $\mu^+\mu^-$ channel ($10$ GeV
to $1$ TeV for $b\bar{b}$), which is the sensitive region explored by
Fermi-LAT.

\begin{figure}[!htb]
\begin{center}
\includegraphics[width=0.47\columnwidth]{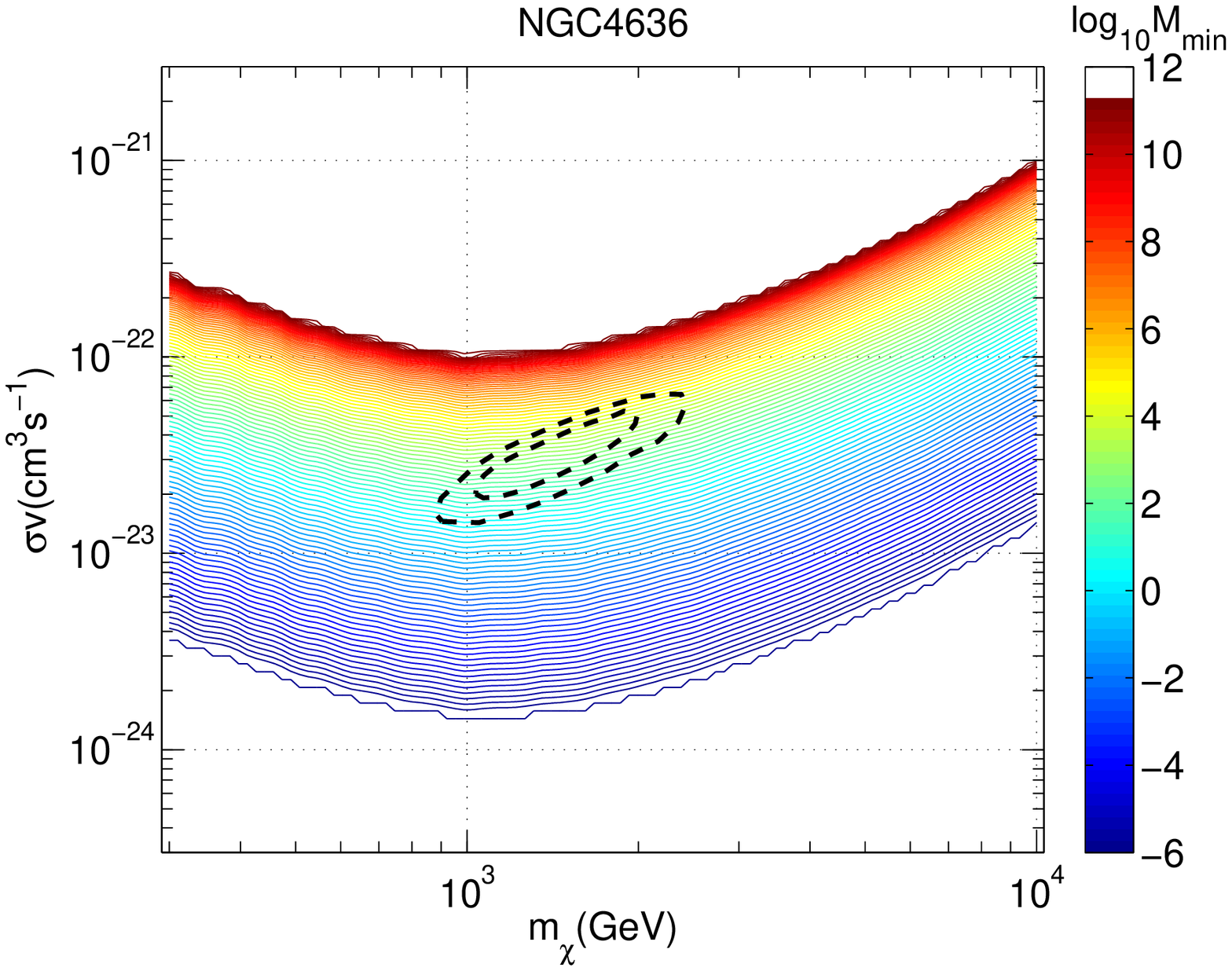}
\includegraphics[width=0.47\columnwidth]{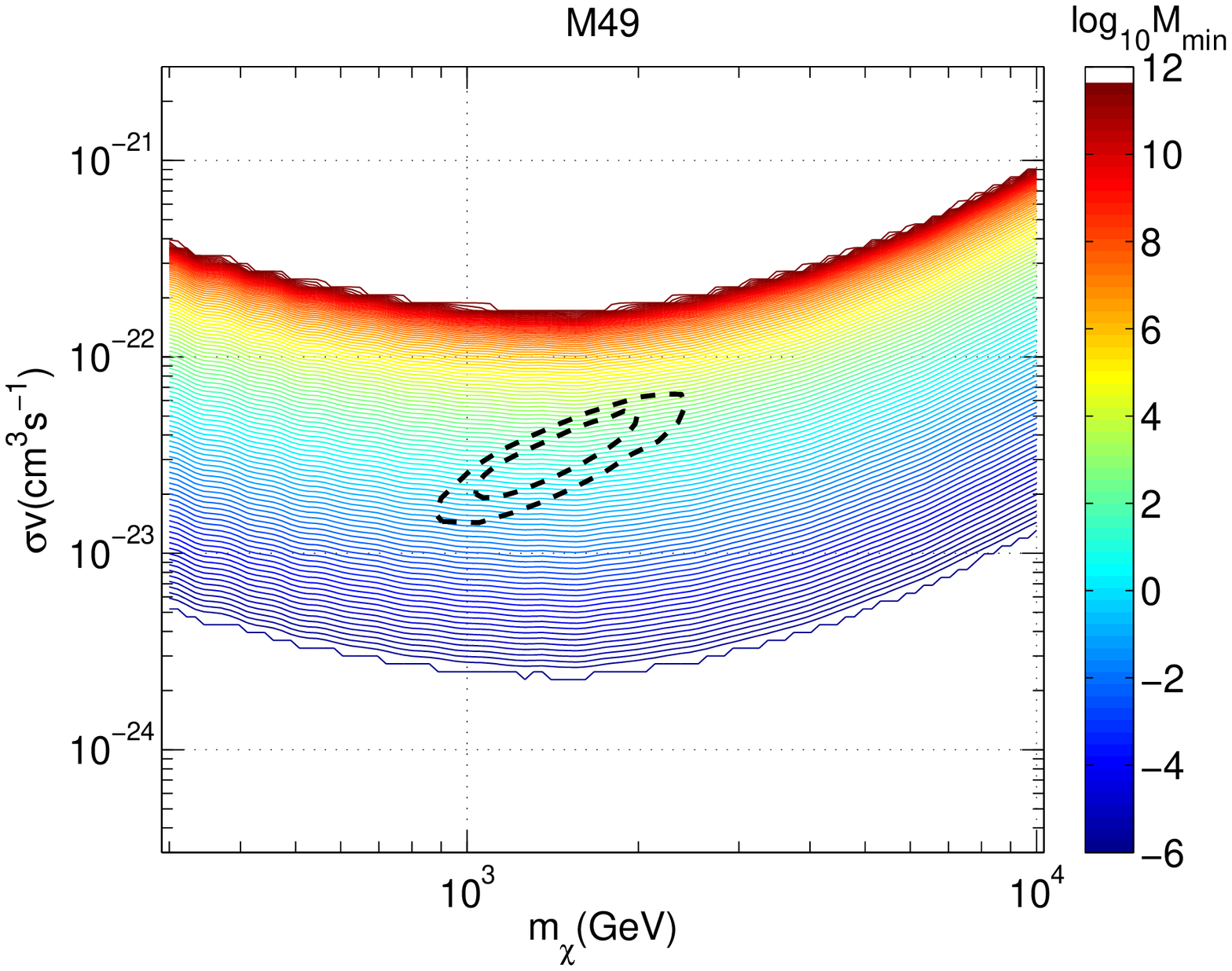}
\includegraphics[width=0.47\columnwidth]{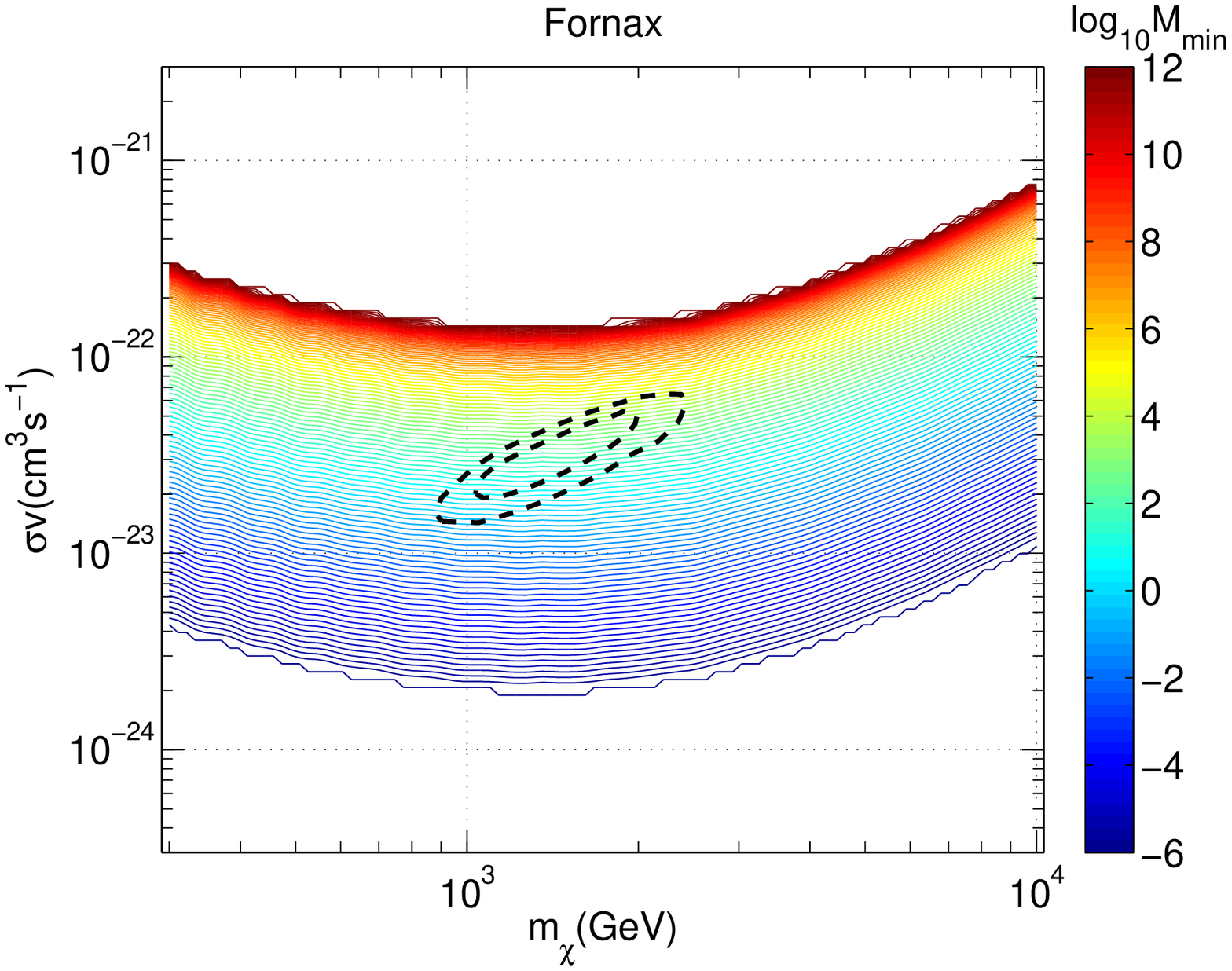}
\includegraphics[width=0.47\columnwidth]{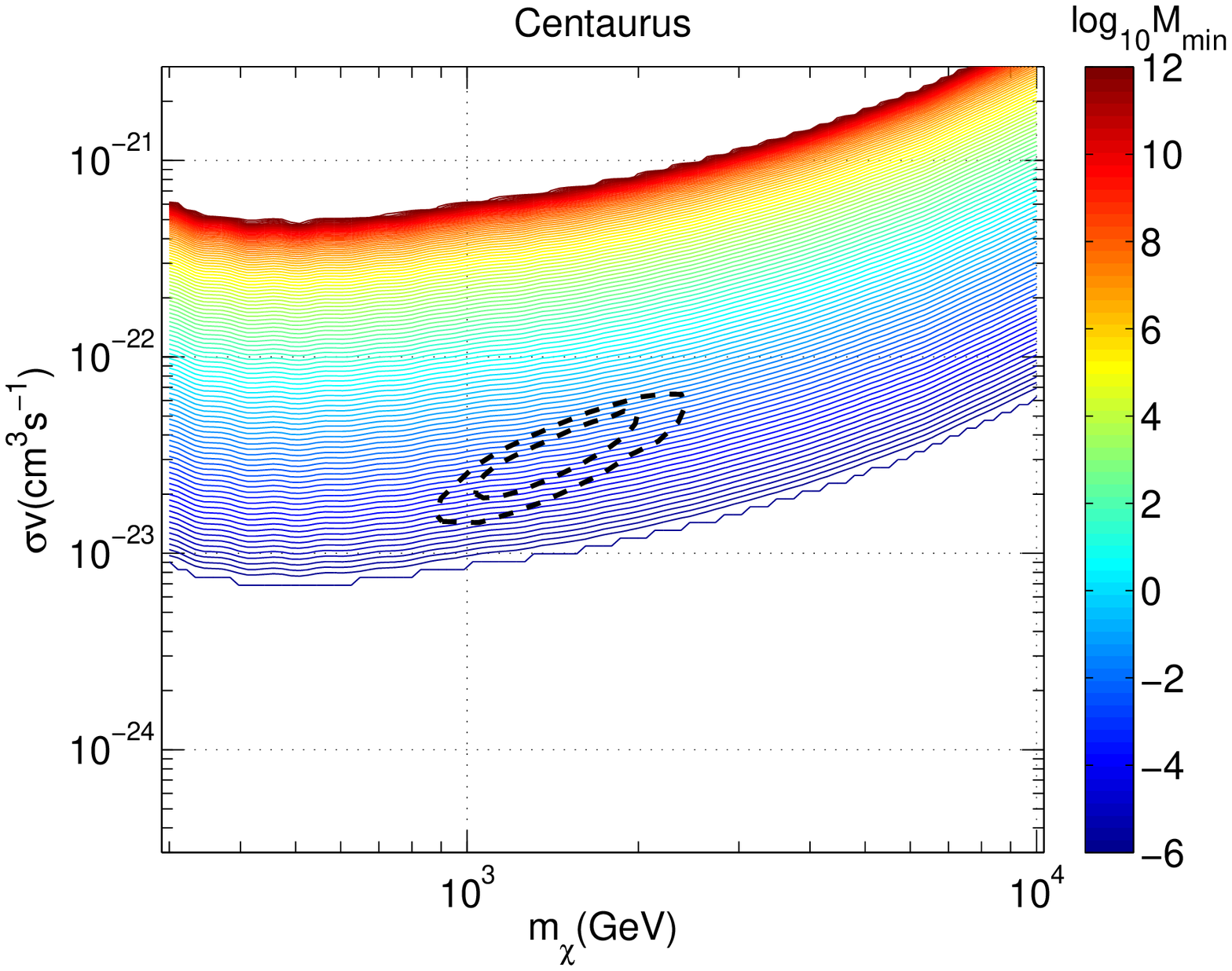}
\includegraphics[width=0.47\columnwidth]{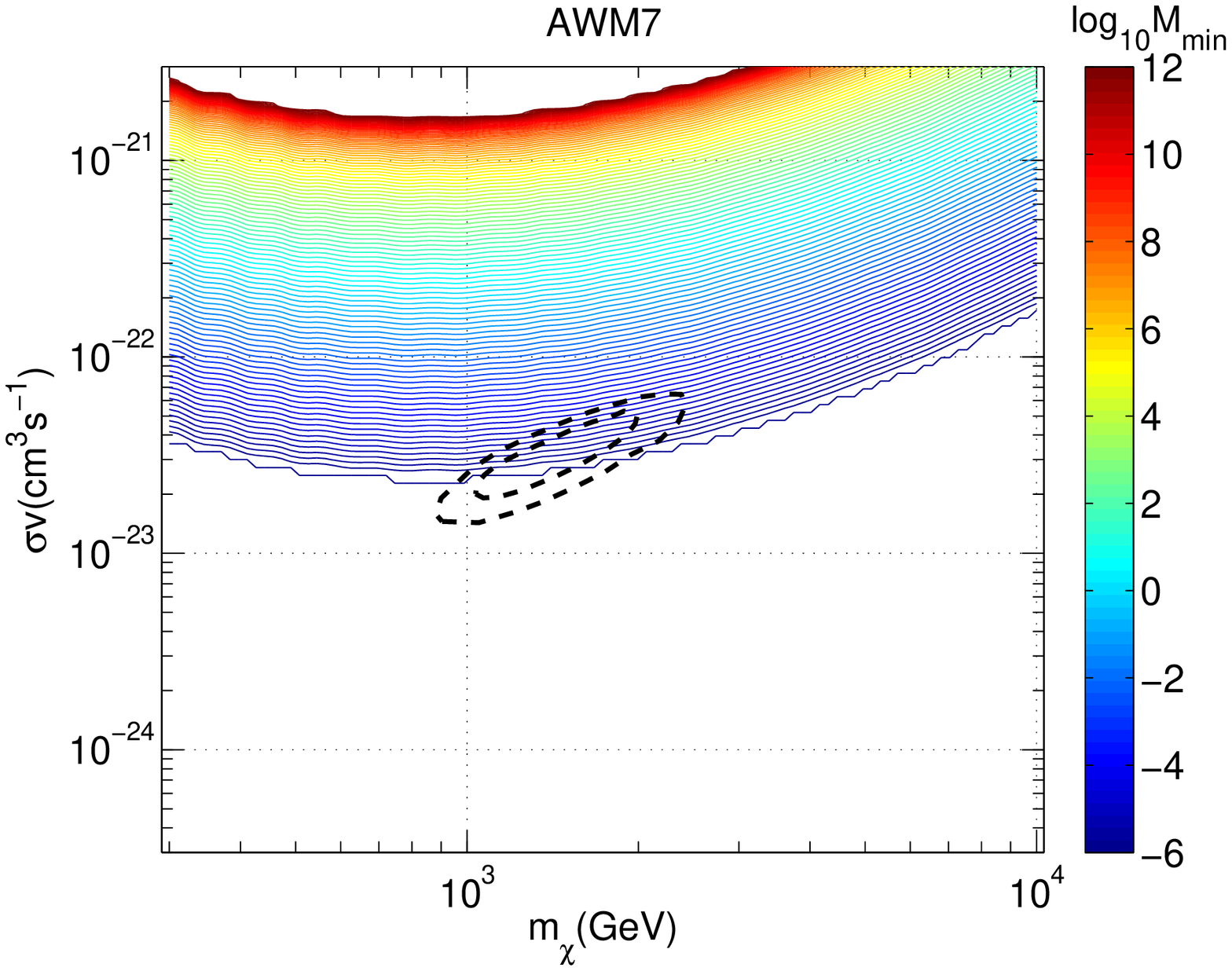}
\includegraphics[width=0.47\columnwidth]{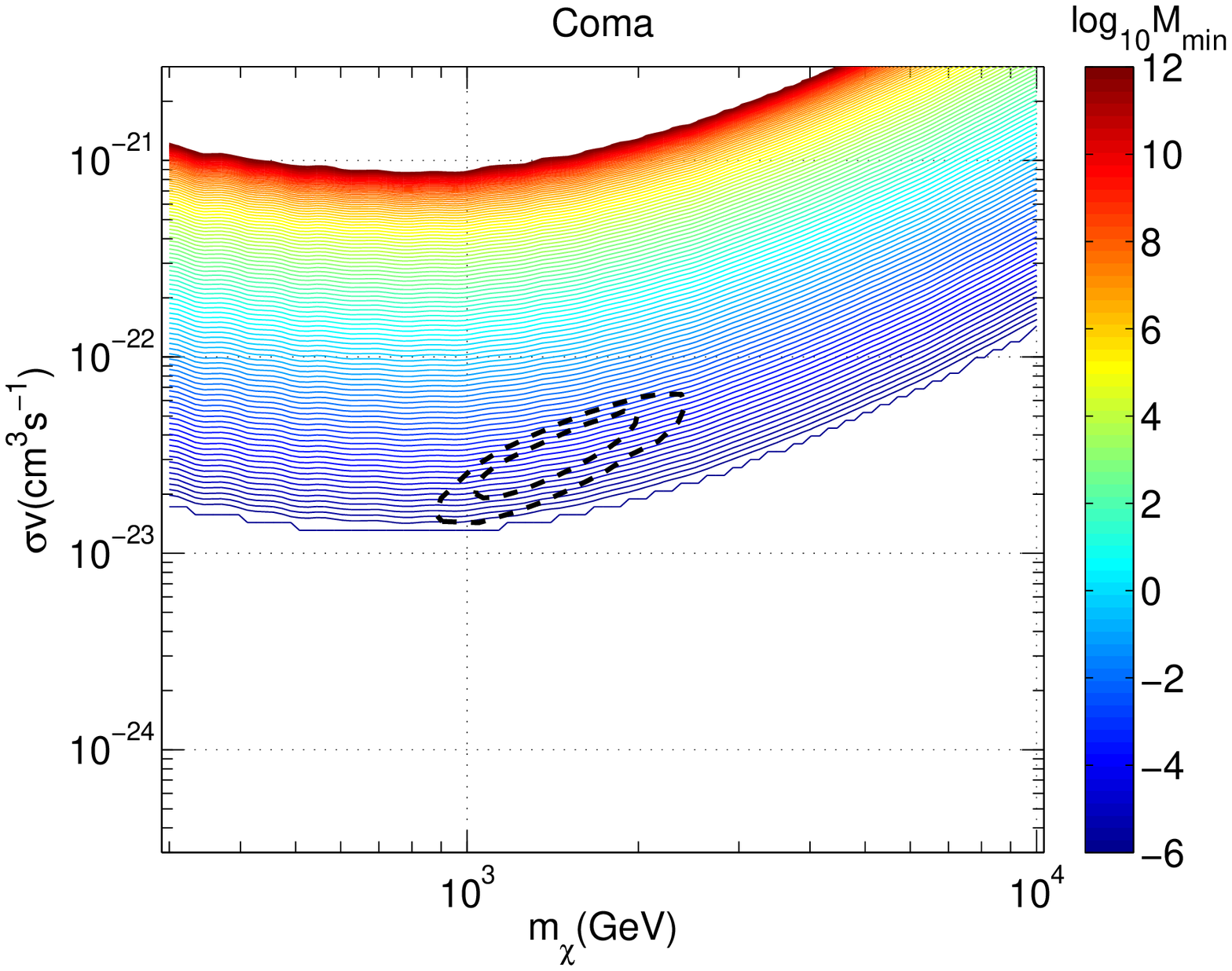}
\caption{Fermi-LAT constraints on the DM model parameters $m_{\chi}$ and
$\langle\sigma v\rangle$ for different minimum halo mass $M_{\rm min}$.
The DM annihilation channel is $\mu^+\mu^-$. Dashed circles are the
$3\sigma$ and $5\sigma$ parameters regions which can fit the
PAMELA/Fermi-LAT/HESS data of the CR positrons/electrons
\cite{2010NuPhB.831..178M}.
\label{fig:mu}}
\end{center}
\end{figure}

\begin{figure}[!htb]
\begin{center}
\includegraphics[width=0.47\columnwidth]{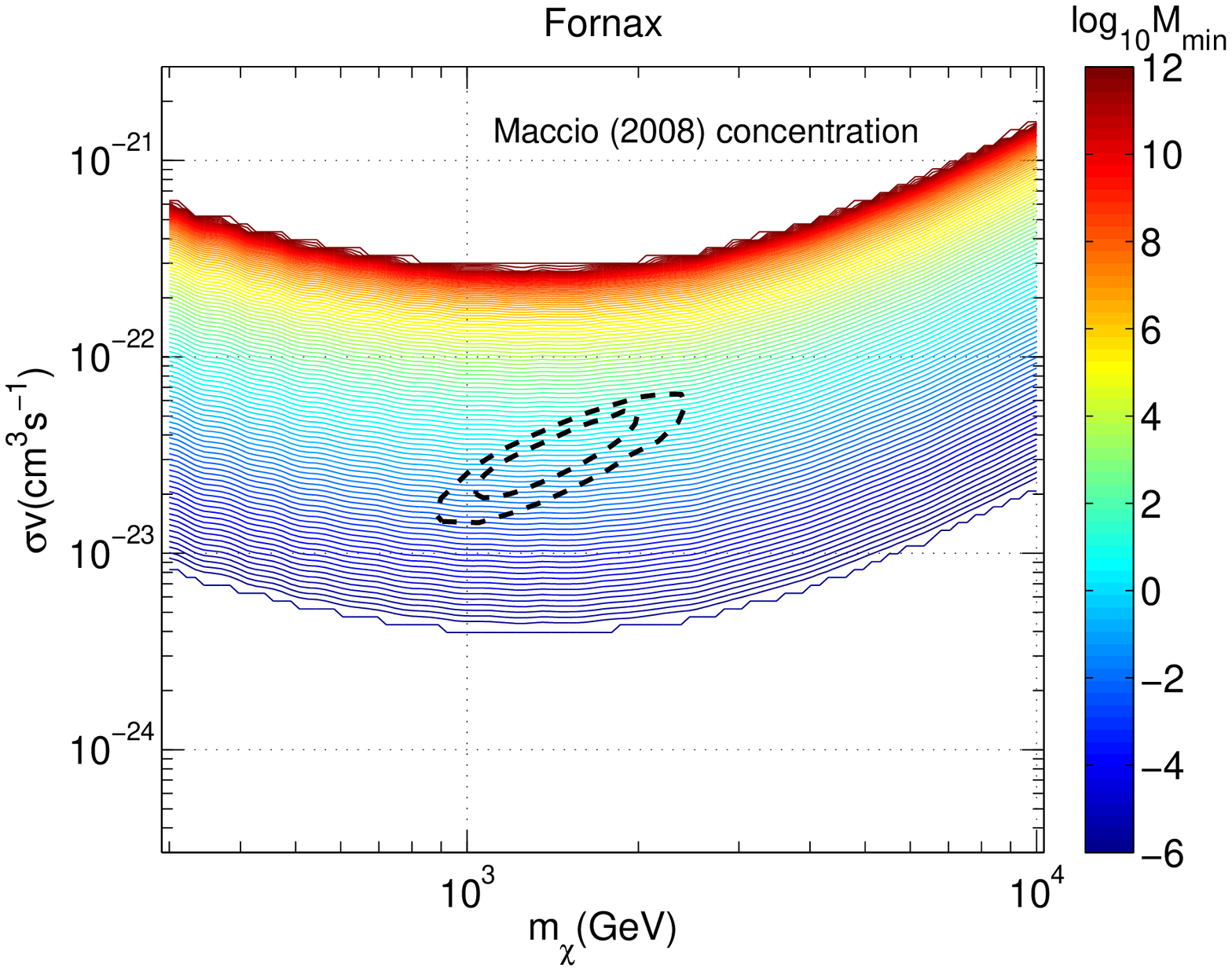}
\includegraphics[width=0.47\columnwidth]{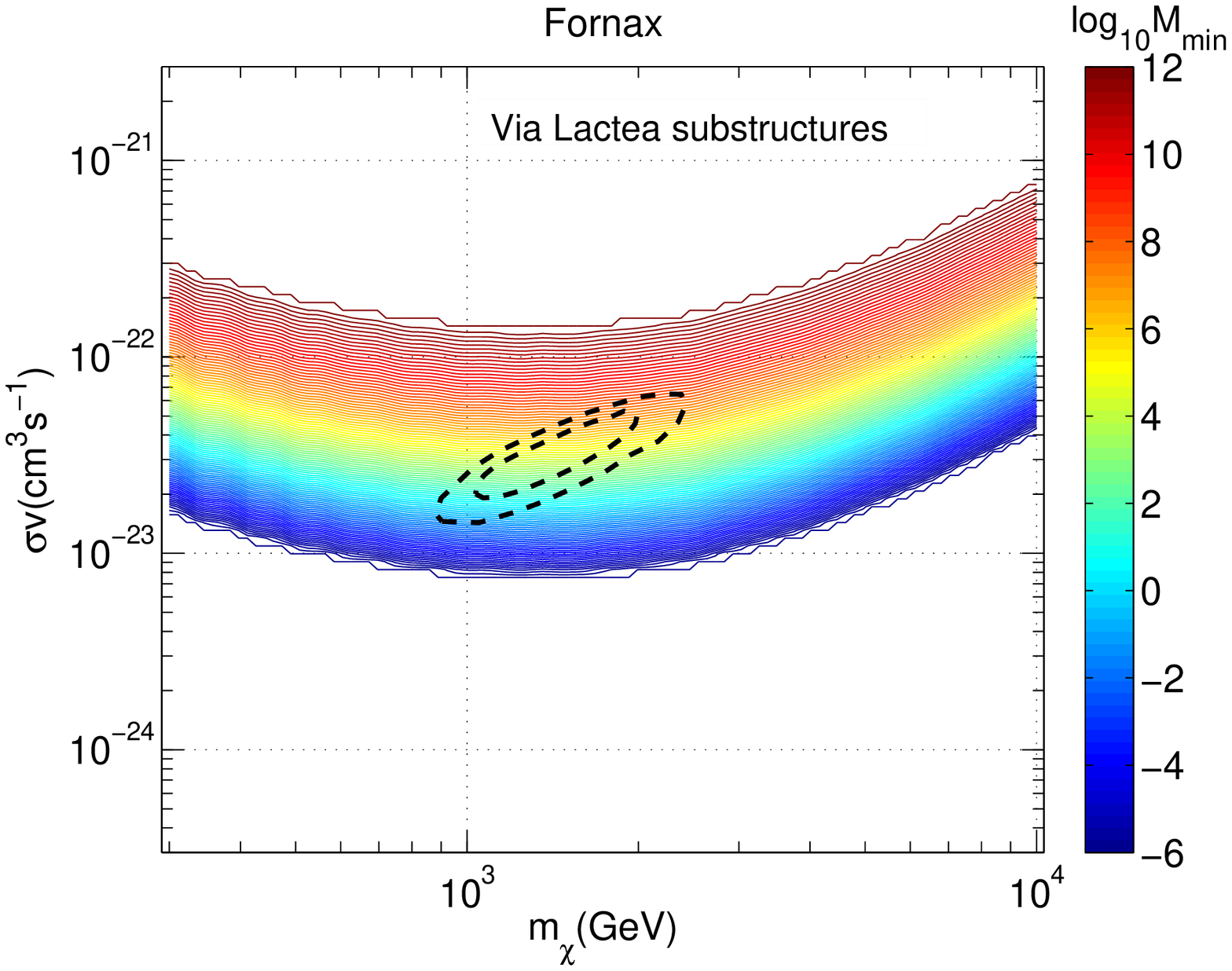}
\caption{Comparison of the constraints for Maccio et al. (2008)
concentration model (left, \cite{2008MNRAS.391.1940M}) and {\it Via
Lactea} substructure model (right, \cite{2008ApJ...686..262K}).
\label{fig:comp}}
\end{center}
\end{figure}

\begin{figure}[!htb]
\begin{center}
\includegraphics[width=0.47\columnwidth]{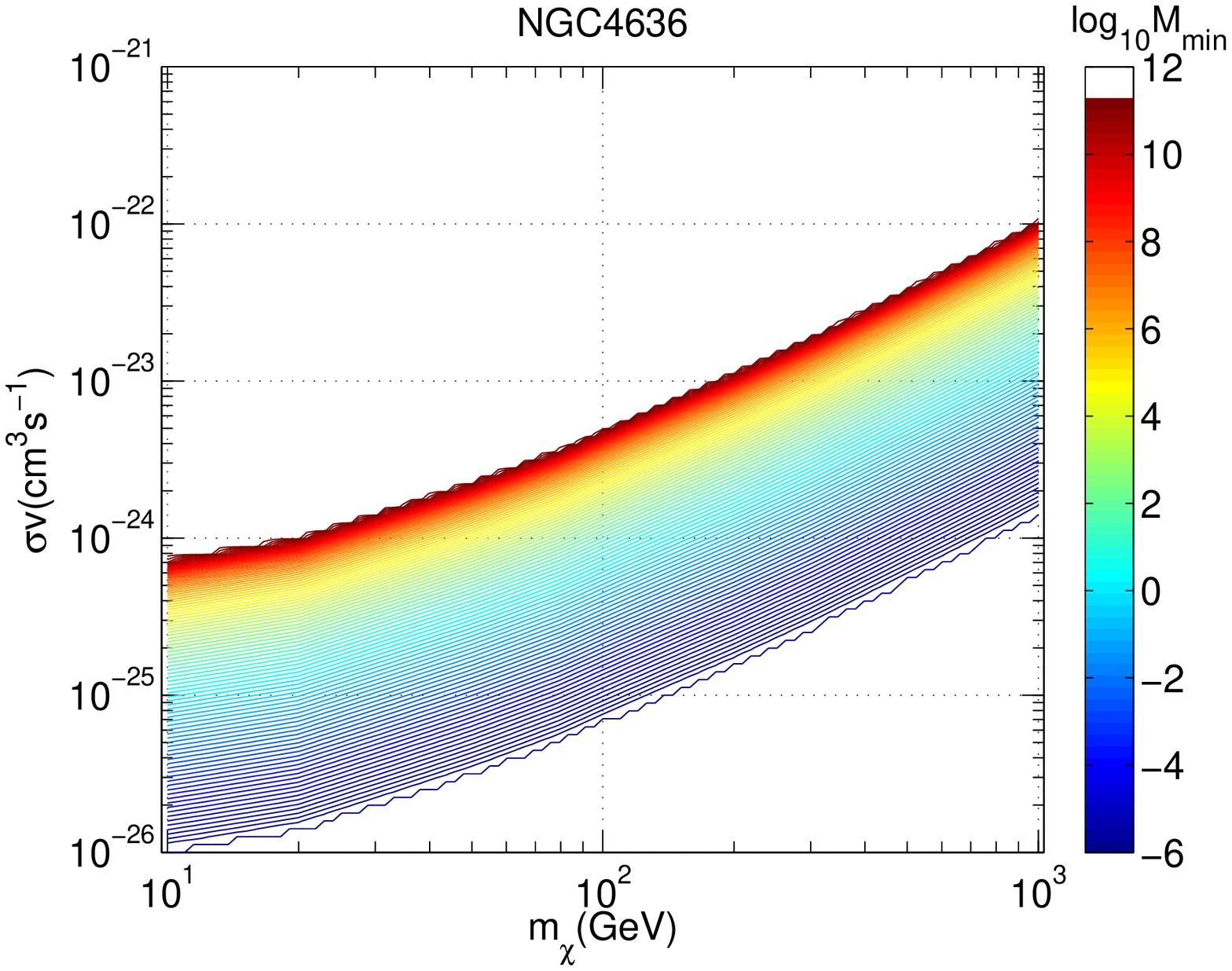}
\includegraphics[width=0.47\columnwidth]{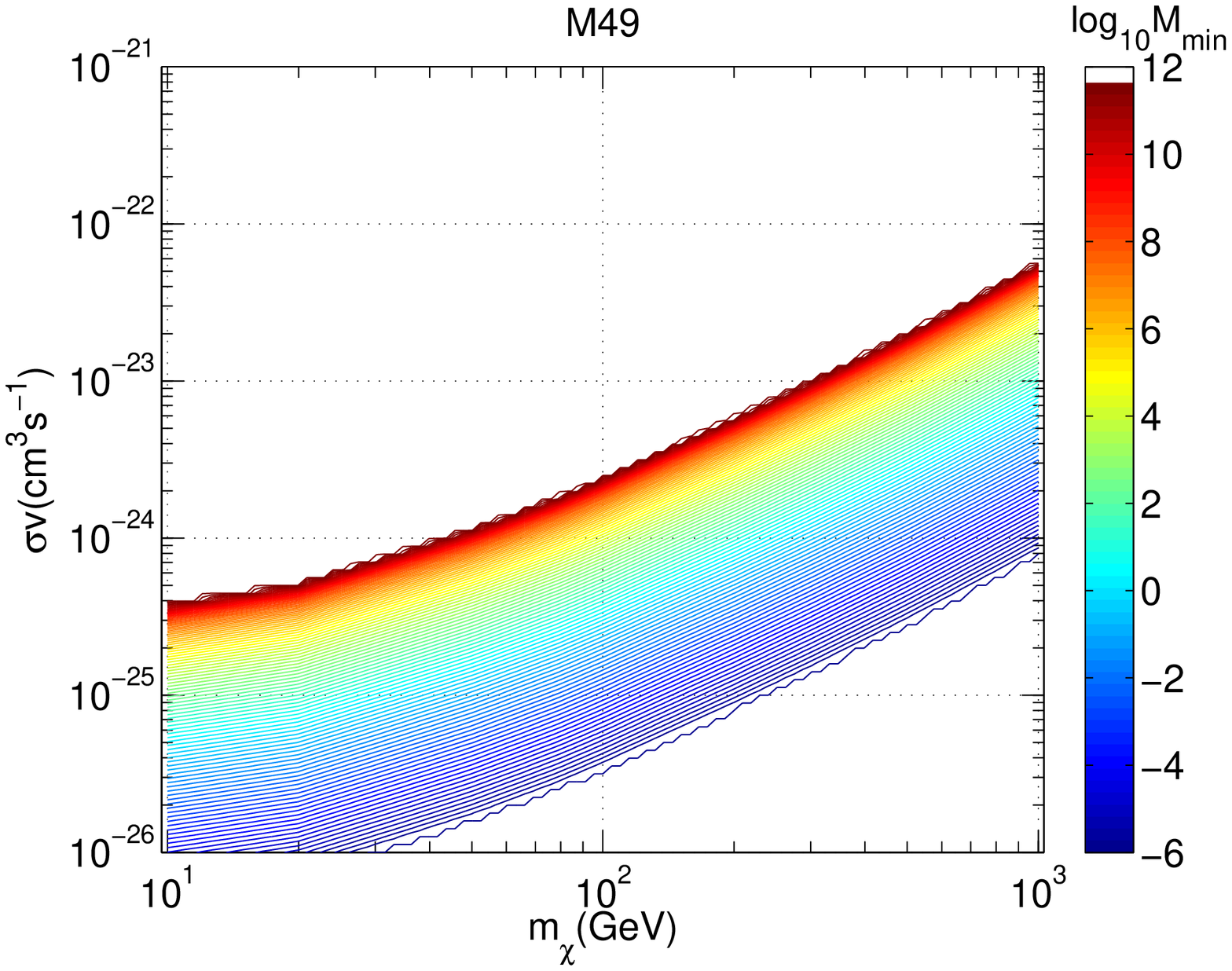}
\includegraphics[width=0.47\columnwidth]{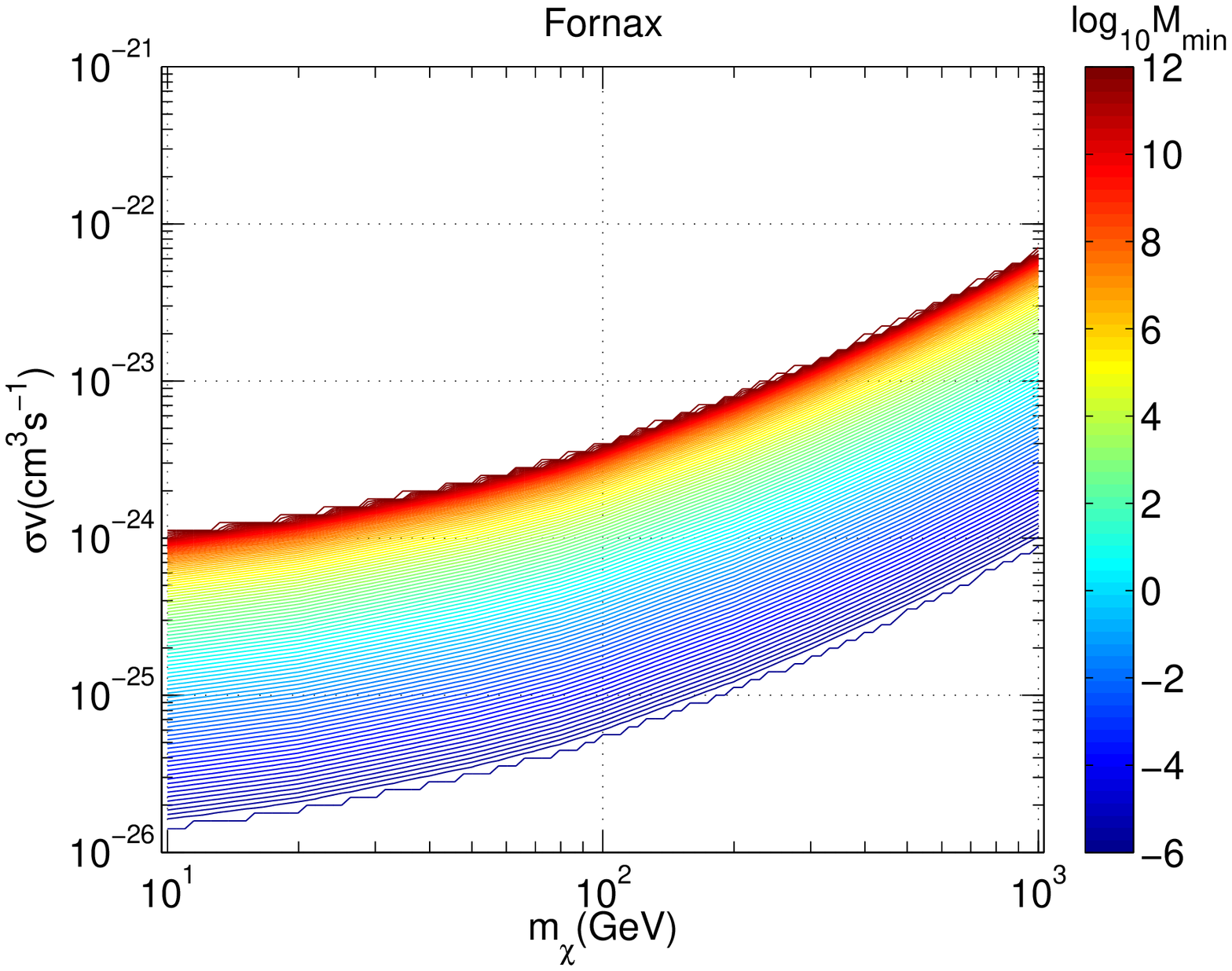}
\includegraphics[width=0.47\columnwidth]{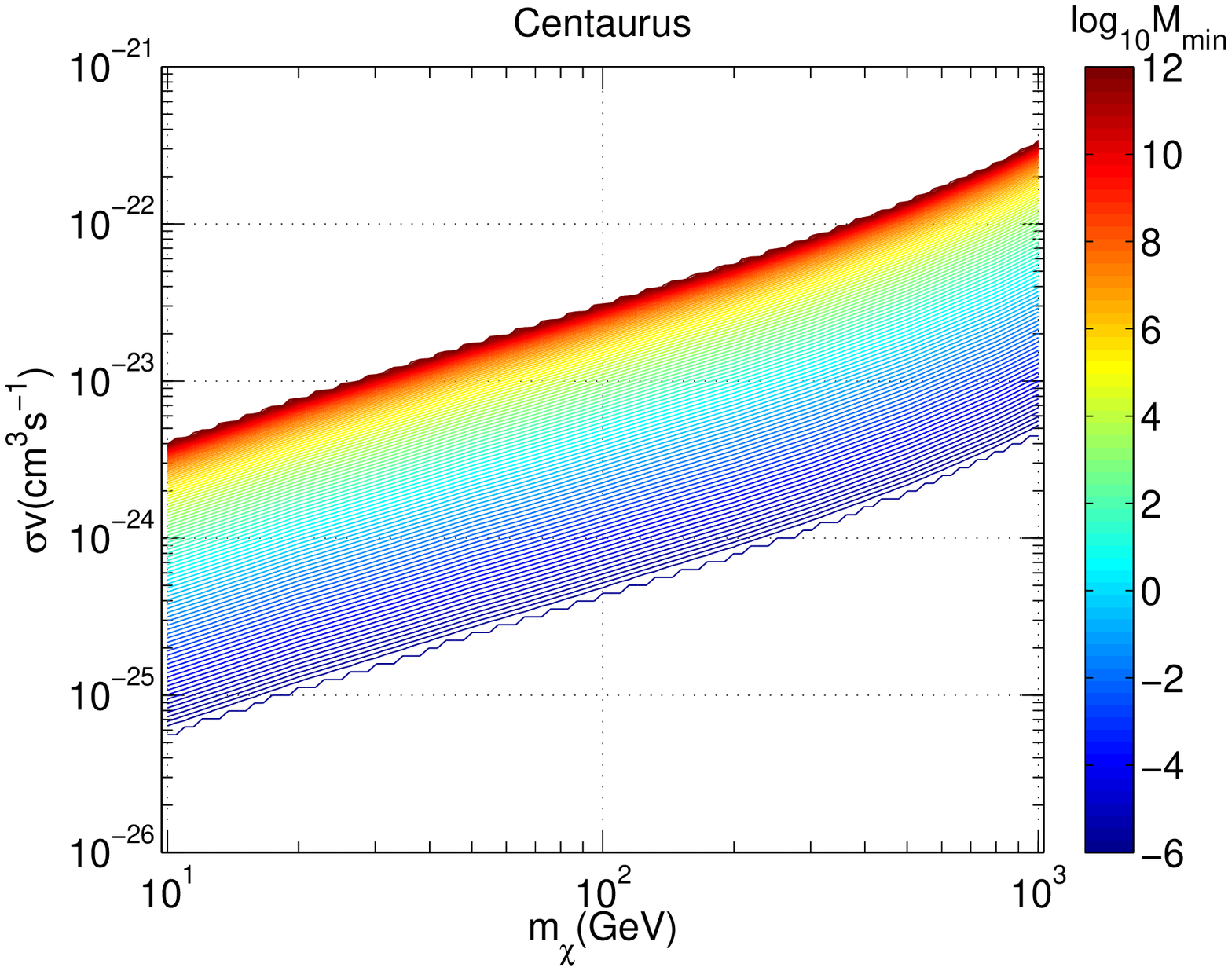}
\includegraphics[width=0.47\columnwidth]{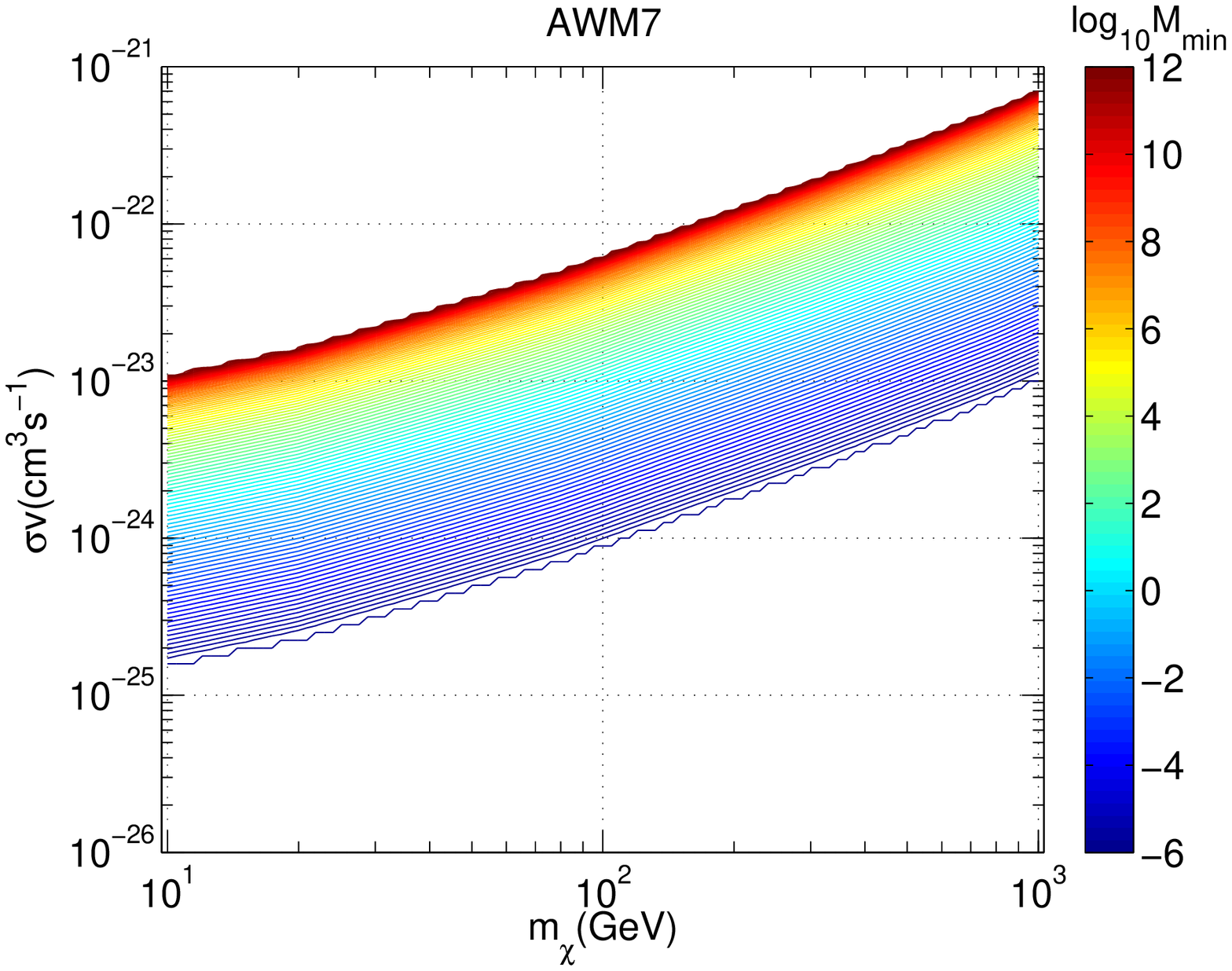}
\includegraphics[width=0.47\columnwidth]{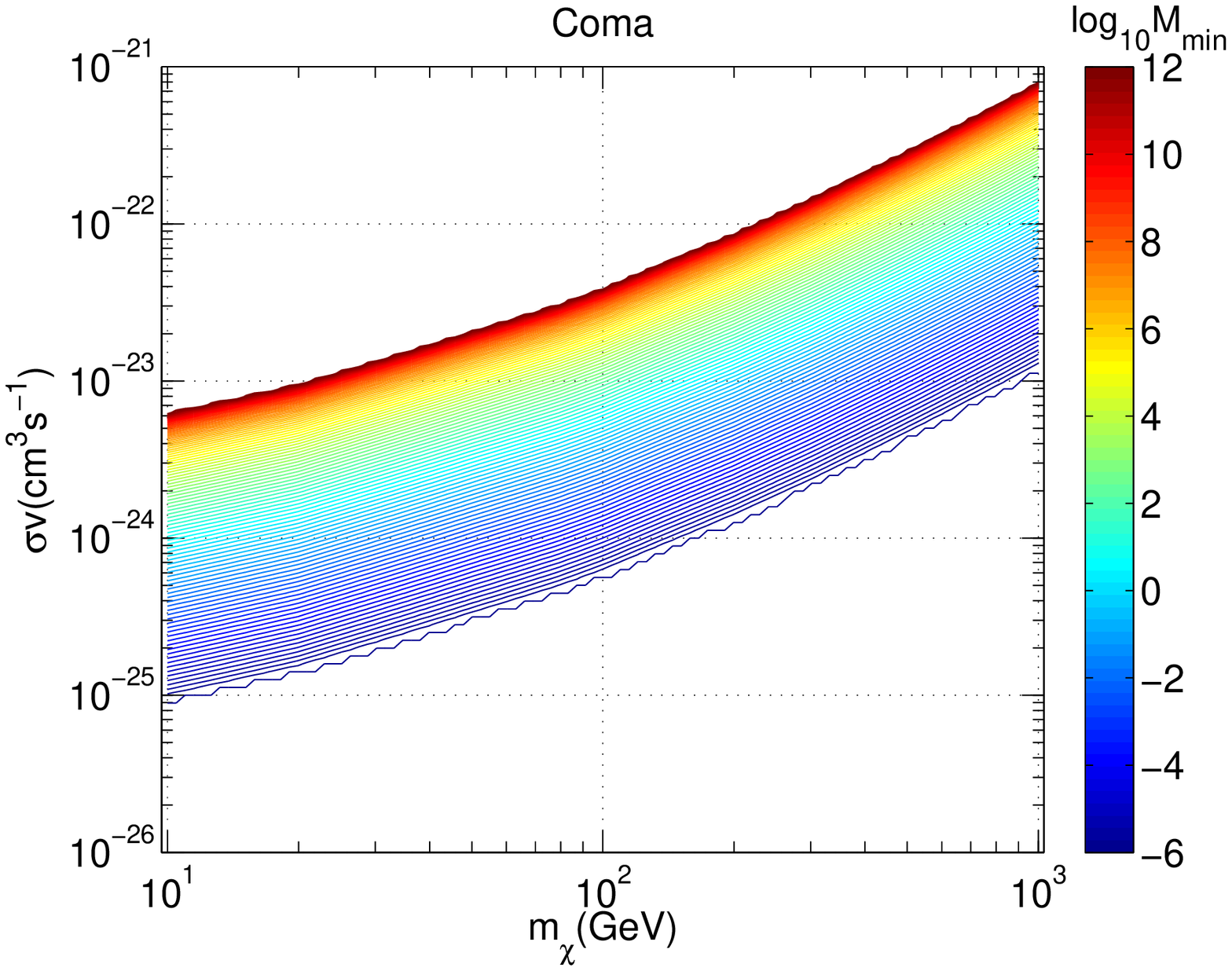}
\caption{The same as Fig. \ref{fig:mu} but for $b\bar{b}$ final state
of DM annihilation.
\label{fig:bb}}
\end{center}
\end{figure}

In \cite{Fermi-LAT:cluster} the Fermi-LAT has derived preliminary
constraints on the $\gamma$-ray fluxes from DM annihilation from
these clusters by assuming the $\mu^+\mu^-$ and $b\bar{b}$ final
states. According to these results, we derive the constraints on
the $m_{\chi}-\langle\sigma v\rangle$ plane. In Figs. \ref{fig:mu}
we show the constraints for the $\mu^+\mu^-$ channel. It is shown
that the constrains of the smooth component (that
$M_{\rm min}=10^{12}M_\odot$) is still very weak. If the substructures
are taken into account the constraints can be stronger by more
than one order of magnitude, depending on the free streaming mass
of DM subhalo. If the minimum mass of DM subhalos can be as low as
$10^{-7}-10^{-6}$ M$_\odot$ as predicted in the supersymmetric DM
scenario, the constraint on the cross section of TeV DM can reach
$\sim 10^{-24}$ cm$^3$ s$^{-1}$. This constraint is actually
powerful enough to explore the DM models which are proposed to
explain the CR lepton excesses. In Fig. \ref{fig:mu} we plot the
favored parameter regions of the DM model to explain the
PAMELA/Fermi-LAT/HESS data of the CR positrons/electrons with
$\mu^+\mu^-$ final state \cite{2010NuPhB.831..178M}. We can see
that for almost all of the 6 clusters the PAMELA/Fermi-LAT/HESS
favored parameter regions can be excluded if $M_{\rm min}$ is down
to $\sim 10^{-6}$ M$_{\odot}$. If DM annihilation is responsible
for the observational positron/electron excesses, it may indicate
that the cutoff of $M_{\rm min}$ should be much larger
\cite{2009PhRvL.103r1302P}. For clusters Fornax and NGC 4636, the
constraint of $M_{\rm min}$ is about $10^{2}-10^{3}$ M$_\odot$ for
the best fit mass and cross section\footnote{Note that here we assume
a constant boost factor of the DM models responsible to the
electron/positron excesses. However, as proposed in the literature
the large boost factor might be due to velocity-dependent annihilation
cross section such as the Sommerfeld effect \cite{2004PhRvL..92c1303H,
2009PhRvD..79a5014A}. In that case the boost factor might be different
in clusters from in the Galaxy. We do not discuss this effect in detail
in this study.}. This constraint is much
stronger than that derived using EGRET upper limit about Virgo
cluster \cite{2009PhRvL.103r1302P}.

This cutoff of the mass of substructures may have important
implication of the particle nature of DM. It gives an estimate of
the free streaming scale of the matter power spectrum as $k<750$
Mpc$^{-1}$, which is not very far from the lower limits given by
Lyman-$\alpha$ power spectrum measurements
\cite{1999ApJ...520....1C,2006ApJS..163...80M}. Compared with the
canonical value expected in CDM picture, it favors a warm massive
DM scenario which may be produced non-thermally in the early
Universe \cite{2001PhRvL..86..954L}. Such nonthermally produced DM
particles have large initial velocity and large free streaming.
Thus the matter power spectrum is suppressed at small scales and
leads to less low mass subhalos \cite{2004PhRvD..69l3521B,
2009PhRvD..80j3502B}.

As a check of the model uncertainties of the halo structure
configuration, we compare the results using Maccio (2008)
concentration-mass relation given in Ref. \cite{2008MNRAS.391.1940M}
\begin{equation}
c_{200}=\frac{3.56}{1+z}\times\left(\frac{M_{200}}{10^{15}
{\rm M}_{\odot}}\right)^{-0.098}.
\end{equation}
For substructures we also use the results from another high-resolution
simulation {\it Via Lactea}. According to Fig. 3 of Ref.
\cite{2008ApJ...686..262K}, the substructure enhancement is simply
extracted to be $B=10\times\left(M_{\rm min}^{-0.048}-M_{\rm max}^{-0.048}
\right)$ with $M_{\rm max}=0.01M_{\rm host}$. Note that there is a host
halo mass dependence of $B$ as given in Ref. \cite{2008ApJ...686..262K}.
However, since the results with $M_{\rm host}$ larger than the galaxy scale
halo were not calibrated in the simulation, we adopt $B(M_{\rm host})$ at
$M_{\rm host}=10^{12}$ M$_{\odot}$ and apply it to the cluster scale halos.
This treatment is also consistent with the above assumption that we
adopt the scaling relation of Eq. (\ref{lumin_sub}) to be the same for
Milky Way halo and cluster halos. The extrapolation to the cluster masses
following the $M_{\rm host}$ dependence of $B$ in Ref.
\cite{2008ApJ...686..262K} would lead to a two times larger boost factor.
The constraints for $\mu^+\mu^-$ channel from Fornax cluster for these
two models are shown in Fig. \ref{fig:comp}. It is shown that for Maccio
(2008) concentration the constraint is about $2$ times weaker than the
benchmark model. And for {\it Via Lactea} substructures the total
substructure enhancement is some weaker than that of {\it Aquarius}
simulation. However, in both of these cases we see that the
PAMELA/Fermi-LAT/HESS favored parameter regions can be constrained,
given $M_{\rm min}\sim 10^{-6}$ M$_{\odot}$. For the best fit mass and
cross section the constraint on $M_{\rm min}$ is about $10^0-10^2$
M$_{\odot}$.

The results for $b\bar{b}$ channel are shown in Fig. \ref{fig:bb}.
We can see that for $m_{\chi}\approx 100$ GeV the most stringent
constraint of $\langle\sigma v\rangle$ can reach the thermal scale
of $3\times 10^{-26}$ cm$^3$ s$^{-1}$ for $M_{\rm min}=10^{-6}$
M$_{\odot}$. The future observation of Fermi-LAT can put a strong
constraint on the DM model even for the thermal scenario.
Note that for $b\bar{b}$ channel only the {\it primary}
emission is assumed when deriving the flux limits. If the IC component
is included the constraints will be stronger for relatively heavy DM
mass ($\sim$TeV). Finally we should point out that $b\bar{b}$ channel
is typically not suitable to explain the lepton excesses observed by
PAMELA/Fermi-LAT/HESS, due to both the constraint from PAMELA
$\bar{p}/p$ data \cite{2009PhRvL.102e1101A,2009NuPhB.813....1C}
and the spectral shape required by lepton excesses
\cite{2009PhRvD..79b3512Y}. Here we include the study of $b\bar{b}$
channel is just to show the power of Fermi-LAT to the supersymmetric-like
DM particles.

\section{Neutrinos}

In this section, we discuss the sensitivity of detecting neutrino
signals from clusters. The cluster could be treated as high energy
neutrino point source, and it is possible to be observed at the
on-going large volume neutrino telescopes, such as IceCube. To
suppress the large atmospheric muon background, the neutrino
telescopes usually detect the upward muons induced by muon neutrinos
through interacting with the matter surrounding the detectors.
Therefore, the south pole based detector IceCube is more suitable to
probe the neutrino sources in the northern hemisphere. For the
sample of clusters in Table \ref{table:sample}, Fornax and Centaurus
locating in the southern hemisphere are not good candidates of
Icecube \footnote{The IceCube + DeepCore has the capability to
search the downward neutrinos, but the angular resolution is fairly
bad \cite{2009NIMPA.602....7F}. It is very difficult to distinguish
the high energy neutrino sources from the high atmospheric neutrino
background without powerful angular resolution.}. AWM 7 with
declination of $41^{\circ}35'$ and Coma with declination of
$27^{\circ}59'$ are suitable for IceCube, but such two clusters are
more distant away from us than other clusters. Taking into account
the location, mass and distance, we find M49 with declination of
$8^{\circ}00'$ is better to be detected than others.

Similar to the {\it primary} $\gamma$-ray flux, the neutrino flux
from DM annihilation in the cluster can also be calculated
according to Eq. (\ref{phi}), by replacing $\left.\frac{{\rm
d}N}{{\rm d}E} \right|_{\gamma}$ with $\left.\frac{{\rm d}N}{{\rm
d}E}\right|_{\nu}$. In the following we will mainly discuss two DM
annihilation channels, $\mu^+\mu^-$ and
$\mu^+\mu^-+\nu_{\mu}\bar{\nu}_{\mu}$. The $b\bar{b}$ channel as
discussed in the previous section will also be mentioned. However,
as we will see below, it gives negligible neutrino signals. We use
the PYTHIA \cite{2006JHEP...05..026S} to simulate the initial
neutrino spectra from decay of annihilation final states. We
further assume the neutrino flavor distribution is $1:1:1$ at the
Earth due to vacuum oscillation during the propagation.

The through-going upward muon rate at the detector can be
calculated as
\begin{equation}
\frac{dN_{\mu}}{dE_{\mu}}=\int d\Omega\int_{E_{\mu}}^{m_\chi}
dE_{\nu_{\mu}} \frac{dN_{\nu_{\mu}}}{dE_{\nu_{\mu}}}\left[
\frac{d\sigma_{\text{CC}}^{\nu p}(E_{\nu_{\mu}},E_{\mu}^0)}{dE_{\mu}^0}
\,n_p+(p\rightarrow n)\right]R(E_\mu)+(\nu\rightarrow\bar{\nu}),
\label{throughmuon}
\end{equation}
where $n_p$ ($n_n$) is the number density of protons (neutrons) in
the matter around the detector, $R(E_{\mu})$ named muon range is the
distance that a muon can travel in matter before its energy drops
below the threshold energy of detector $E_{\rm th}$, which is given by
\begin{equation}
R(E_\mu) = \frac{1}{\rho \beta} \ln {\frac{ \alpha + \beta E_{\mu}}{
\alpha + \beta E_{\rm th}} } ,
\end{equation}
with $\alpha , \beta$ the parameters describing the energy loss of
muons as $dE_{\mu}/dx=-\alpha-\beta E_{\mu}$.

The main background for high energy neutrino detection is the
atmospheric neutrinos. The atmospheric neutrino flux decreases
rapidly as energy increasing. We use a parametrization of atmospheric
neutrino flux \cite{2009PhRvD..80d3514E} which describes the results
of Ref. \cite{2007PhRvD..75d3006H} as
\begin{equation}
\frac{{\rm d} N_{\nu}}{{\rm d} E_\nu {\rm d} \Omega}=N_0
E_\nu^{-2.74}\left(\frac{0.018}{1+0.024 E_\nu |\cos
\theta|}+\frac{0.0069}{1+0.00139 E_\nu |\cos \theta|}\right),
\end{equation}
where $N_0=1.95\times 10^{17}$ ($1.35\times 10^{17}$)
GeV$^{-1}$km$^{-2}$yr$^{-1}$sr$^{-1}$ for $\nu_\mu$
($\bar{\nu}_\mu$) respectively, $\theta$ is the zenith angle.

IceCube could have an angular resolution $\sim 1^\circ$, which is
effective to suppress the diffuse atmospheric neutrino background.
Notice the cluster is not an exact point source, we utilize an
angular resolution of 3.0$^\circ$ (1.5$^\circ$) for M49
(AWM47/Coma) cluster. The number of atmospheric neutrinos for
3$^\circ$ resolution angle is $\sim 4$ times larger than it for
1.5$^\circ$ cone.

\begin{figure}[!htb]
\begin{center}
\includegraphics[width=0.47\columnwidth]{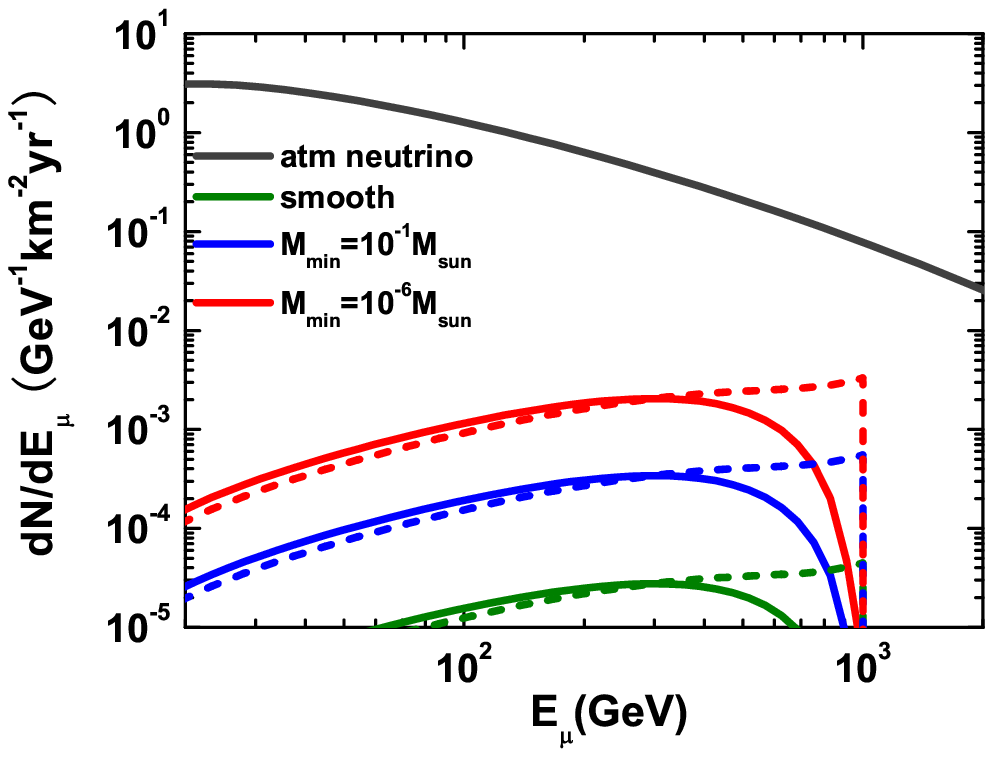}
\includegraphics[width=0.47\columnwidth]{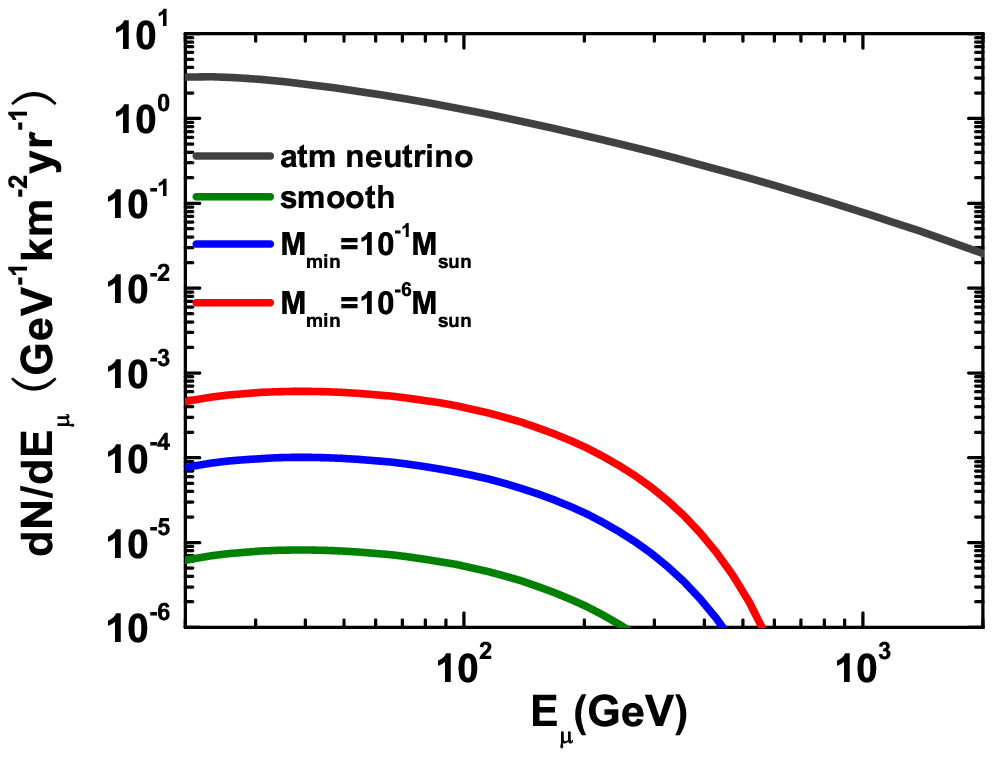}
\caption{Muon flux induced by neutrinos from DM annihilation in M49
cluster, for muon (left) and $b\bar{b}$ (right) final states
respectively. The mass of DM is $1$ TeV and cross section is $10^{-23}$
cm$^{3}$ s$^{-1}$. The results of smooth halo (green) and smooth halo
together with subhalos with two values of $M_{\rm min}$, $10^{-1}$ M$_{\odot}$
(blue) and $10^{-6}$ M$_{\odot}$ (red) are shown. In the left panel, the
solid lines denote $\mu^+ \mu^-$ channel, and dashed lines denote
$\mu^+ \mu^-$ plus $\nu_\mu\bar{\nu}_\mu$ with equal branching ratios.
Also shown is muon flux induced by atmospheric muon neutrino background
in 3$^\circ$ cone.
\label{fig:M49spec}}
\end{center}
\end{figure}

In Fig. \ref{fig:M49spec}, we show the through-going muon flux
induced by DM annihilation in M49 cluster. Similar as in Fig.
\ref{fig:spec} we adopt $m_\chi=1$ TeV, $\langle\sigma
v\rangle=10^{-23}$ cm$^3$ s$^{-1}$, and annihilation channels are
$\mu^+\mu^-$ (left solid), $\mu^+\mu^-+\nu_\mu \bar{\nu}_\mu$
($B_{\mu}=B_{\nu}=0.5$, left dashed) and $b \bar{b}$ (right). For
$b\bar{b}$ channel, the neutrinos are produced through decay of
hadrons induced by $b \bar{b}$ hadronization. It is shown that the
muon spectrum of such channel is very soft and difficult to detect
given the high level of background. The case for $\mu^+\mu^-$
channel seems better but the signal is still very weak. Even for
$M_{\rm min}=10^{-6}$ M$_{\odot}$, the muon flux from DM
annihilation is $\sim 100$ times smaller than the background in
energy range $(200,\,800)$ GeV. Only for $\mu^+\mu^-+\nu_\mu
\bar{\nu}_\mu$ the situation is better. The reason is that
monochromatic neutrino spectrum is harder than other channels, and
is easier to be detected. Compared with Fig. \ref{fig:spec}, it is
not strange to see that the neutrino detection sensitivity would
be much weaker than the $\gamma$-ray detection.

\begin{figure}[!htb]
\begin{center}
\includegraphics[width=0.47\columnwidth]{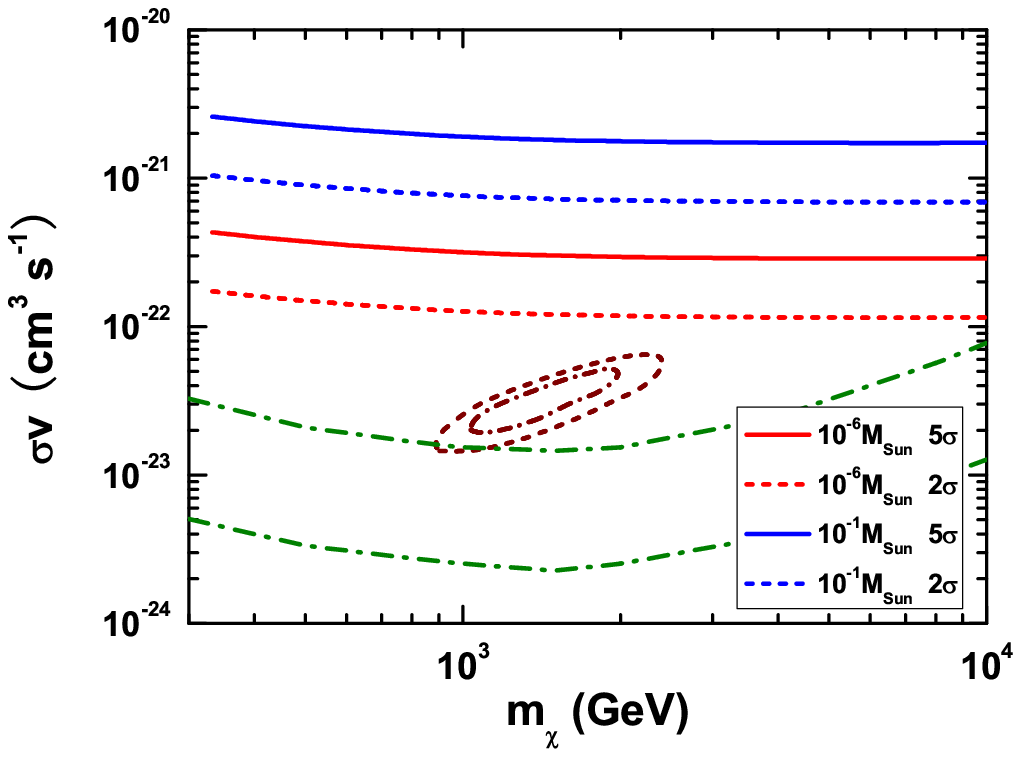}
\includegraphics[width=0.47\columnwidth]{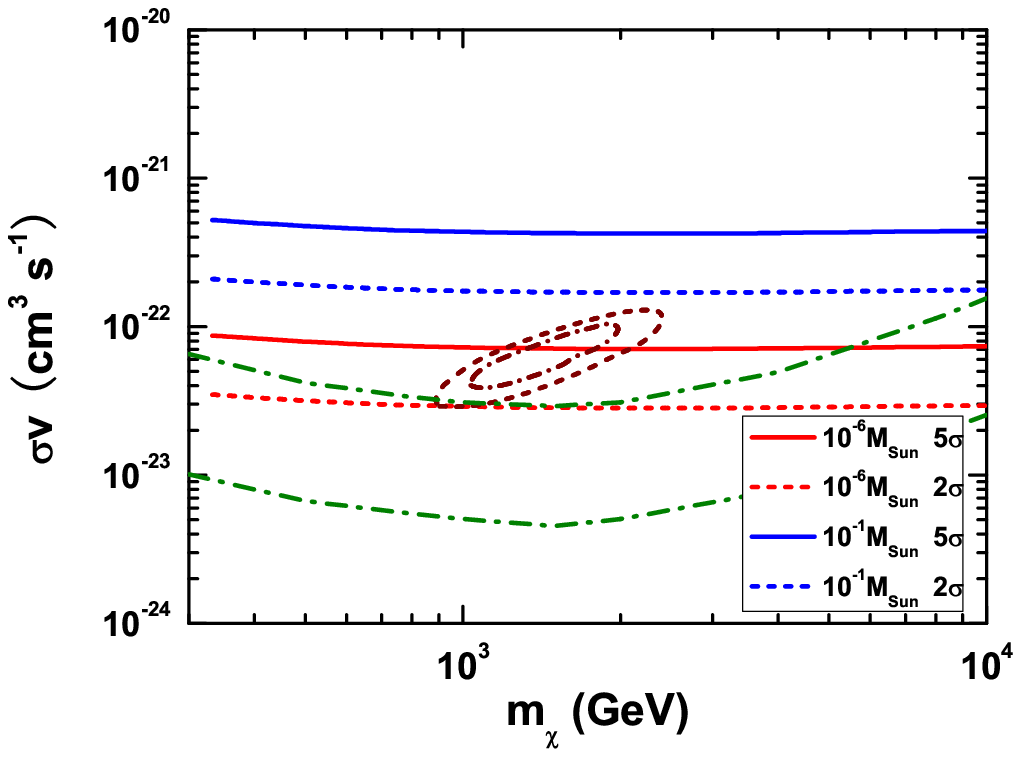}
\caption{$\langle\sigma v\rangle$ required to discover neutrinos from DM
annihilation in M49 cluster as a function of DM mass. The left panel
is for $\mu^+ \mu^-$ channel, and the right panel is for $\mu^+\mu^- +
\nu_\mu \bar{\nu}_\mu$ with branching ratios $B_{\mu}=B_{\nu}=0.5$.
Dashed circles are the $3\sigma$ and $5\sigma$ parameters regions which
can fit the PAMELA/Fermi-LAT/HESS data of the CR positrons/electrons
\cite{2010NuPhB.831..178M}. The dot-dashed curves represent the
$\gamma$-ray constraints for $M_{\rm min}=10^{-1}$ and $10^{-6}$
M$_{\odot}$ respectively (see Fig. \ref{fig:mu}). The circles and
$\gamma$-ray constraints in the right panel are scaled
upwards by a factor $2$ due to the branching ratio $B_{\mu}=0.5$.
\label{fig:M49nu}}
\end{center}
\end{figure}

\begin{figure}[!htb]
\begin{center}
\includegraphics[width=0.47\columnwidth]{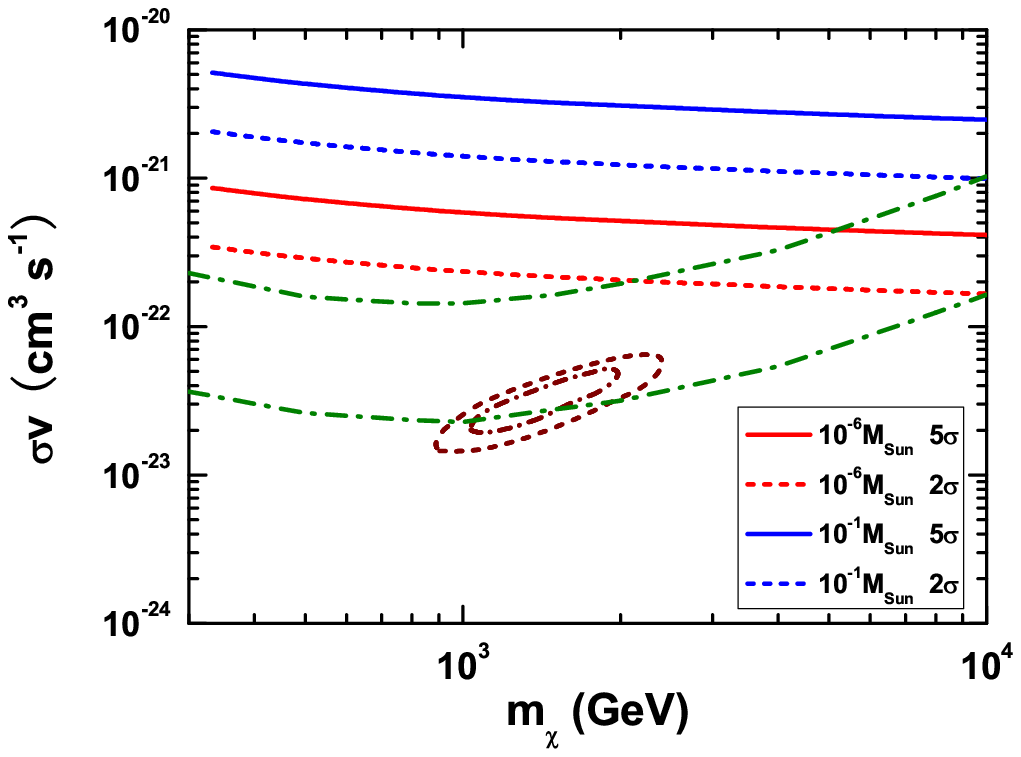}
\includegraphics[width=0.47\columnwidth]{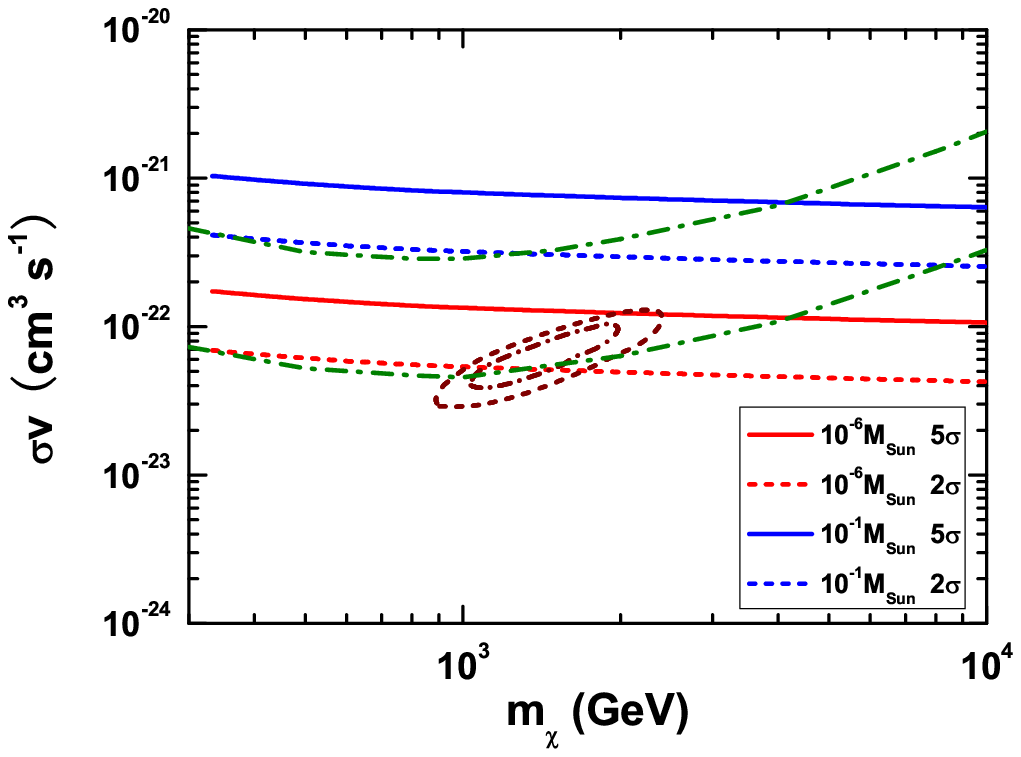}
\caption{The same as Fig. \ref{fig:M49nu} but for AWM7 cluster.
\label{fig:AWM7nu}}
\end{center}
\end{figure}

\begin{figure}[!htb]
\begin{center}
\includegraphics[width=0.47\columnwidth]{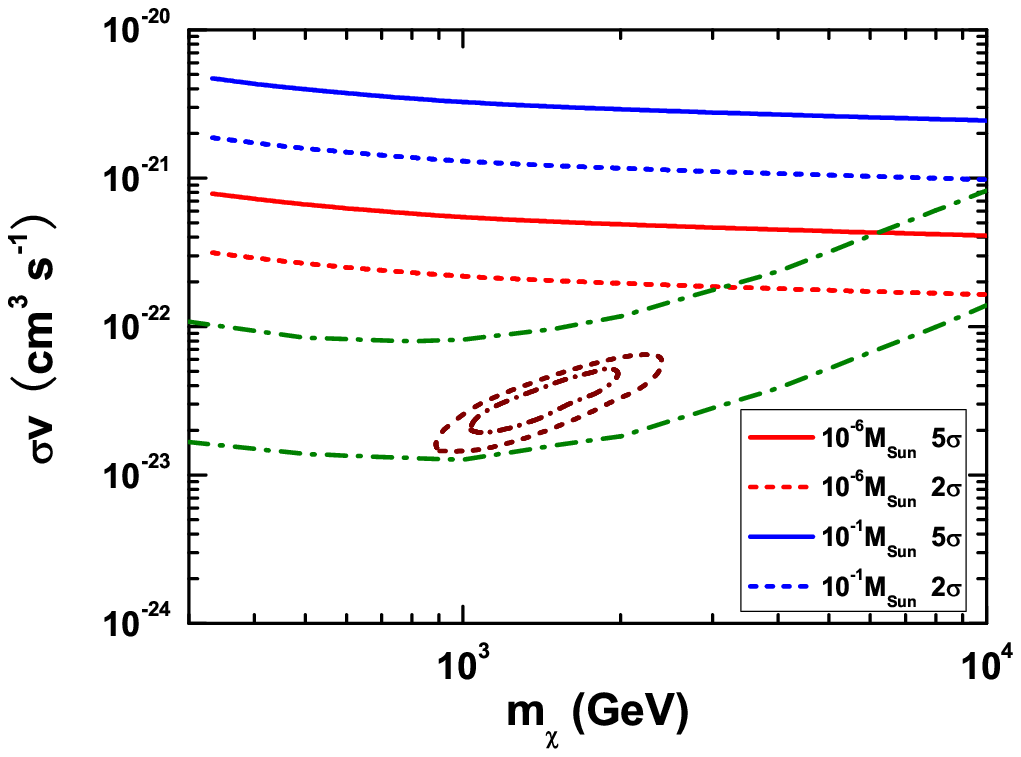}
\includegraphics[width=0.47\columnwidth]{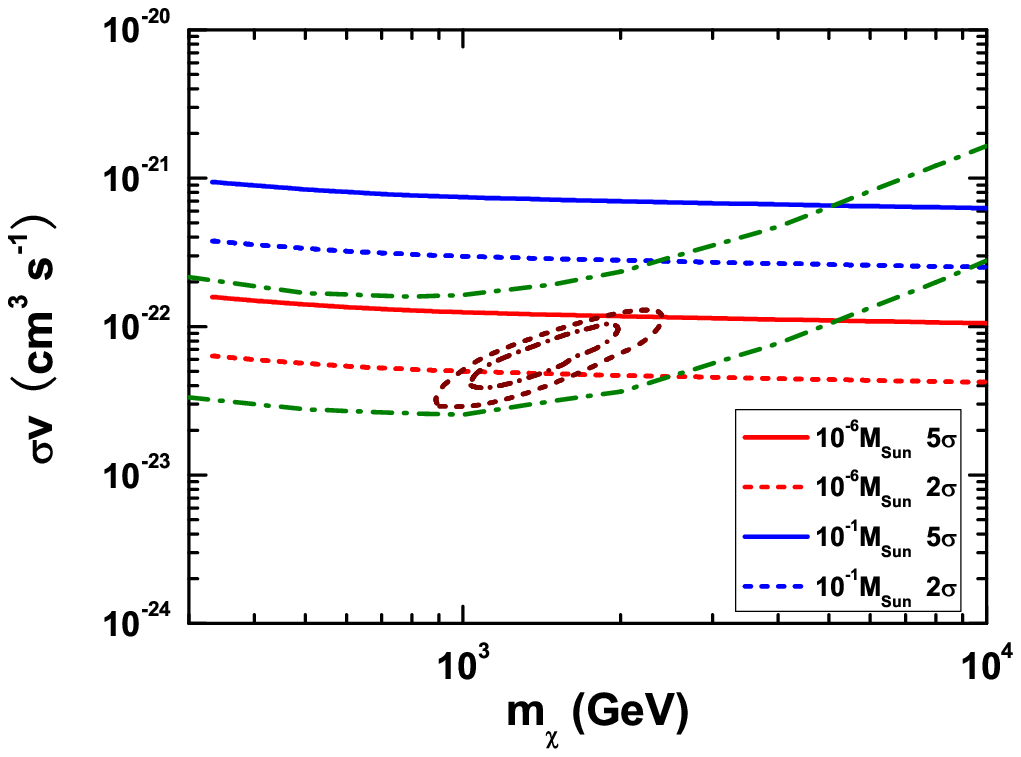}
\caption{The same as Fig. \ref{fig:M49nu} but for Coma cluster.
\label{fig:Comanu}}
\end{center}
\end{figure}

The total muon event rate in a specific energy bin at the detector is
\begin{equation}
N=\int {\rm d}E_\mu \frac{{\rm d}N_\mu}{{\rm d}E_\mu} A_{\rm eff}(E_{\mu},
\theta) \Delta T,
\end{equation}
where $A_{\rm eff}$ is the effective muon detecting area taken
from Ref. \cite{2009APh....31..437G}, $\Delta T$ is the operation time
which is set as $10$ years here. We take the threshold energy to be
$E_{\rm th}\sim$40 GeV, and assume the energy resolution is $\Delta
\log_{10}E=0.35$ \cite{2009NIMPA.602....7F}.

In Fig. \ref{fig:M49nu} we give the IceCube sensitivity of detecting
neutrinos from DM annihilations in M49 for $10$-yr exposure. The left
and right panels show the results for annihilation channels $\mu^+\mu^-$
and $\mu^+\mu^- + \nu_\mu \bar{\nu}_\mu$ respectively. For $\mu^+\mu^-$
channel and $M_{\rm min}=10^{-6}$ M$_{\odot}$, a $5\sigma$ detection
requires $\langle\sigma v\rangle$ as large as $O(10^{-22})$ cm$^3$
s$^{-1}$. We can see such a cross section is much larger than that
required to explain PAMELA/Fermi-LAT/HESS data. For the channel to
equal $\mu^+\mu^- + \nu_\mu \bar{\nu}_\mu$, the sensitivity is $\sim 10$
times better than $\mu^+\mu^-$ channel. Because the neutrino-nucleon
cross section and muon range are approximately proportional to $E_\nu$,
while the atmospheric neutrino background decreases as $\sim E_\nu^{-3}$,
the neutrino telescope is more powerful to detect the high energy neutrinos.
For the same reason, the detector is more sensitive to explore heavy DM.
It is shown that if DM annihilation products have a large branching ratio
to $\nu \bar{\nu}$, IceCube could reach the parameter space to explain
PAMELA/Fermi-LAT/HESS results.

The sensitivities for other two clusters, AWM 7 and Coma, are shown in
Figs. \ref{fig:AWM7nu} and \ref{fig:Comanu} respectively. These two
clusters are much more distant from us than M49. Note for these two
clusters we take $1.5^\circ$ cone, which is enough to include all of
the cluster, to count the atmospheric background. Due to a much lower
level of background and larger masses of AWM 7 and Coma, the sensitivities
are only several tens percent weaker than M49. Because the declinations
of AWM 7 and Coma are larger than M49 which is close to the horizon,
it would be more effective to reject the CR muon background and could
provide clearer detection of signals.

Compared with the $\gamma$-ray sensitivity discussed in Sec. II, the
sensitivity of neutrino detection is relatively poor. If we employ the
$\gamma$-ray constraint of DM annihilation in clusters (e.g., $M_{\rm min}
\approx 10^3$ M$_{\odot}$ for PAMELA/Fermi-LAT/HESS best fit parameters),
there would be little chance to detect the neutrino signals. Only for very
heavy DM ($m_\chi\sim 10$ TeV), the sensitivities of the two ways are
comparable.

\section{Summary and discussion}

In this work we investigate the $\gamma$-ray and neutrino emission
from DM annihilation in clusters of galaxies. A sample of several
nearby clusters is considered. Both the annihilations in the host
halo and substructures are taken into account. For the
annihilation luminosity of substructures we adopt the result of
{\it Aquarius} simulation, and scale it from Milky Way like halo
to the cluster like halo. There is also a component of unresolved
substructures which is not seen due to the limit of resolution of
numerical simulation. For the contribution from the unresolved
substructures we adopt an extrapolation of the luminosity-mass
relation fitted from the resolved subhalos in the simulation. The
minimum mass of the subhalo $M_{\rm min}$ is left to be a free
parameter. The value of $M_{\rm min}$ may catch the information
about free streaming length, and may reflect the generation
history of the DM particle.

For the $\gamma$-ray emission we consider two typical annihilation
channels, quark final states $b\bar{b}$ and lepton final states
$\mu^+\mu^-$. Both the {\it primary} component of photons from
hadron decay and final state radiation, and the {\it secondary} component which
is produced by the IC scattering of DM-induced
electrons/positrons and CMB field are taken into account. The
Fermi-LAT upper limits of the $\gamma$-ray emission from these
clusters are employed to constrain the $m_{\chi}-\langle\sigma
v\rangle$ parameter plane. The constraints depend on the value of
$M_{\rm min}$. For $M_{\rm min}$ down to $\sim 10^{-6}$
M$_{\odot}$, which is expected for the structure formation of
neutralino-like CDM picture, the constraints are $\sim 50$ times
stronger than the case of smooth halo only. Typically for
$m_{\chi}=1$ TeV and $M_{\rm min}=10^{-6}$ M$_{\odot}$ the
strongest constraint on cross section from the cluster sample is
$10^{-24}$ ($10^{-24}$) cm$^3$ s$^{-1}$ for $\mu^+\mu^-$
($b\bar{b}$).

It is of great interest for the $\mu^+\mu^-$ channel which is
proposed to explain the positron and electron excesses reported by
PAMELA, Fermi-LAT and HESS
\cite{2009PhRvL.103c1103B,2010NuPhB.831..178M}. A very large
annihilation cross section (or boost factor) is needed to explain
the data. It is shown that the Fermi-LAT observations about
$\gamma$-rays from galaxy clusters can strongly constrain the
model parameters. If we fix the mass and cross section to the
values explaining the $e^{\pm}$ excesses, the minimum mass of
substructures $M_{\rm min}$ is constrained to be larger than
$10^2-10^3$ M$_{\odot}$. Such a large value of halo mass means a
very large free streaming length of DM particle. It may indicate
the nature of DM particles is warm instead of cold
\cite{2001PhRvL..86..954L,2009PhRvL.103r1302P}.

Finally we calculate the sensitivity of detecting neutrinos from
these clusters by the IceCube detector. It is shown to detect
neutrinos would be much more difficult than $\gamma$-rays. For
$b\bar{b}$ final state the sensitivity is extremely poor due to
the neutrino spectrum from $b\bar{b}$ hadronization is soft and
suffers from a very high atmospheric background. The case becomes
better for $\mu^+\mu^-$ final state. However, the signal is still
very weak. For example, even for $M_{\rm min}=10^{-6}$
M$_{\odot}$, a $5\sigma$ detection with $\sim 10$-yr exposure
requires $\langle\sigma v\rangle$ as large as $O(10^{-22})$ cm$^3$
s$^{-1}$. This result does not have enough capability to explore
the PAMELA/Fermi-LAT/HESS favored parameter region. If we consider
the model with a large fraction of line neutrino emission the
detectability would be much better (e.g. improved by an order of
magnitude). However, compared with $\gamma$-rays the constraint
from neutrinos is still some weaker. Only for very heavy DM
($m_\chi\sim 10$ TeV), the sensitivity of neutrino detection can
be comparable with $\gamma$-rays.

\acknowledgments

This work was supported in part by the Natural Sciences Foundation
of China under the grant Nos. 10773011, 10775001, 10635030, the
973 project under grant 2010CB833000, and the trans-century fund
of Chinese Ministry of Education.

\bibliography{refs}
\bibliographystyle{apsrev}

\end{document}